\documentclass[twocolumn,trackchanges]{aastex701}
\usepackage{amsmath,amstext}
\usepackage{apjfonts}
\usepackage[figure,figure*]{hypcap}
\usepackage{ulem}
\usepackage{xspace}
\usepackage{xcolor}
\usepackage{lineno}
\usepackage{xfrac}
\usepackage{graphicx}
\usepackage{multimedia}
\usepackage{amssymb}
\usepackage{array,booktabs}
\usepackage{mathptmx}
\usepackage{float}
\usepackage{multirow}
\newcommand{\mathsc}[1]{\text{\normalfont\texttt{#1}}}
\newcommand{\dd}{\mathrm{d}}
\newcommand{\vect}[1]{\boldsymbol{\mathbf{#1}}}

\newcommand{\s}{\,{\rm s}}
\newcommand{\cm}{\,{\rm cm}}
\newcommand{\g}{\,{\rm g}}
\newcommand{\km}{\,{\rm km}}
\newcommand{\erg}{\,{\rm erg}}
\newcommand{\mbh}{\,{M_{\rm BH}}}
\newcommand{\rg}{\,{r_{\rm g}}}
\newcommand{\rb}{\,{R_{\rm B}}}
\newcommand{\td}{\,{t_{\rm disk}}}
\newcommand{\msun}{\,{M_{\odot}}}

\graphicspath{{./}{figures/}}

\shorttitle{The Life and Death of Stars That Capture Primordial Black Holes}
\shortauthors{Gottlieb, Cantiello, Norton, Van Tilburg \& Kleban}

\begin{document}
\title{
The Life and Death of Stars That Capture Primordial Black Holes}

\correspondingauthor{Ore Gottlieb}
\email{ogot@mit.edu}

\author[0000-0003-3115-2456]{Ore Gottlieb}
\affil{Department of Physics and Kavli Institute for Astrophysics and Space Research, Massachusetts Institute of Technology, Cambridge, MA 02139, USA}
\email{ogot@mit.edu}

\author[0000-0002-8171-8596]{Matteo Cantiello} 
\affil{Center for Computational Astrophysics, Flatiron Institute, 162 5th Avenue, New York, NY 10010, USA} 
\affil{Department of Astrophysical Sciences, Princeton University, Princeton, NJ 08544, USA}
\email{mcantiello@flatironinstitute.org}

\author[0000-0002-4382-9885]{Cameron Norton} 
\email{cen8858@nyu.edu}
\affil{Center for Cosmology and Particle Physics, Department of Physics, New York University, New York, NY 10003 USA}

\author[0000-0001-7085-6128]{Ken Van Tilburg}
\email{kenvt@stanford.edu}
\affil{Leinweber Institute for Theoretical Physics, Department of Physics, Stanford University, Stanford, CA 94305, USA}

\author[0000-0002-1889-2487]{Matthew Kleban} 
\affil{Center for Cosmology and Particle Physics, Department of Physics, New York University, New York, NY 10003 USA}
\email{kleban@nyu.edu}

\begin{abstract}
Primordial black holes (PBHs) in the asteroid mass window ($10^{17}$--$10^{23}\,{\rm g}$) remain viable dark matter candidates and can be captured by stars. We develop the first global framework for the evolution of stars that capture PBHs, combining analytic calculations, stellar evolution models, 3D general-relativistic magnetohydrodynamic simulations, and Monte Carlo population synthesis. We find that the fate of these systems bifurcates: PBHs that form an accretion disk before consuming the host drive explosive disruption, whereas PBHs captured too late or growing too slowly consume the star quietly. Capture is dominated by three-body interactions with planetary or stellar companions. For a solar-type host with a Jupiter analog, inspiral within a main-sequence lifetime requires $M_{\rm BH}^{\rm crit}\gtrsim 10^{22}\,{\rm g}$, while lighter PBHs generally require tighter companions. Once deposited at the center, the PBH grows through inefficient quasi-spherical Bondi accretion; if it reaches the angular-momentum threshold before consuming the host, the inflow circularizes into a disk. Our Monte Carlo calculations yield sizable quiet-consumption and explosive-disruption populations, with final PBH masses $M_{\rm BH}\sim0.01$--$1\,M_\odot$ and disk-forming PBH spins $a_\ast\approx0.8$. Disk formation is the point of no return: disk winds and relativistic jets of $\sim10^{45}$--$10^{50}\,{\rm erg\,s^{-1}}$ disrupt the star within minutes. The resulting transients may include a $\sim$day-long UV/blue signal, radio afterglow, and, if the jet escapes, an X-ray-flash/low-luminosity gamma-ray-burst (XRF/llGRB) signal. For an $\mathcal{O}(1)$ PBH dark matter fraction and optimistic capture assumptions, the event rate can reach that of llGRBs. The low-mass, high-spin remnants offer a complementary PBH probe and possible source for subsolar BH mergers.
\end{abstract}

\section{Introduction} \label{sec:intro}

Primordial black holes (PBHs) are of broad interest in cosmology and astrophysics as potential dark matter (DM) candidates and as probes of early-Universe physics. They can arise from the gravitational collapse of large-amplitude density perturbations generated by inflation \citep{Carr1975, Ivanov1994, GarciaBellido1996,Clesse2015}, from cosmological phase transitions \citep{Hawking1982,Jedamzik1997}, or from other early-Universe processes (see \citealt{Khlopov2010,Green2021,Carr2026} for reviews). Depending on their mass spectrum and abundance, PBHs could account for a non-negligible fraction of the DM \citep{Escriva2024}, contribute to the merger rate observed by LIGO--Virgo--KAGRA \citep{Bird2016,Sasaki2016}, and seed early halos or supermassive black holes (BHs) at high redshift \citep{Bean2002,Bromm2003}. These possibilities motivate continued efforts to constrain the PBH mass spectrum and abundance using astrophysical observations. While a variety of cosmological and astrophysical probes place strong limits on PBHs at very small and very large masses, substantial gaps remain in the intermediate regime \citep[e.g.,][]{Gorton2024,Carr2026}.

PBHs with mass $ \mbh \lesssim 10^{15}\,\g$ are excluded by Hawking evaporation \citep{Carr1974,Hawking1975,Page1976}. PBHs with slightly larger masses, $ \mbh \sim 10^{15}-10^{17}\,\g $ are tightly constrained by the absence of excess gamma rays and cosmic rays produced by Hawking radiation from long-lived PBHs \citep[e.g.,][]{Boudaud2019,DeRocco2019,Coogan2021,Ray2021,Korwar2023}. At the other end of the spectrum, microlensing surveys \citep[e.g.,][]{Green2016,Niikura2019a,Niikura2019b,Alcock2021,Tisserand2007,Blaineau2022,Mroz2024,Smyth2020}, wide-binary disruption \citep[e.g.,][]{Yoo2004,Quinn2009,Monroy2014}, and the dynamical heating of stellar systems tightly constrain PBHs at stellar masses and above \citep[e.g.,][]{Lacey1985,Brandt2016,Koushiappas2017}. Between these low- and high-PBH mass constraints, there remains a broad mass window spanning asteroid to sublunar masses, $ \mbh \sim 10^{17}-10^{23}\,\g$, where existing bounds weaken, and observational baselines are limited, allowing PBHs to comprise a non-negligible fraction of DM \citep[e.g.,][]{Carr2010,Carr2016,Carr2021,Carr2026}.

Stars provide a qualitatively different and complementary probe of this remaining parameter space. If PBHs constitute a significant fraction of the DM, their number density in galactic environments would be vastly larger than that of stellar-origin BHs, implying frequent close encounters with stars, making capture by stellar systems a plausible outcome \citep[e.g.,][]{Montero2019,Lehmann2022,Oncins2022,Esser2023}. Once a PBH is captured, it migrates to the stellar core, where it accretes from the surrounding material and influences the host star’s evolution in ways that are inaccessible in diffuse environments. The resulting object, a ``Hawking star'' \citep{Hawking1971,Bellinger2023}, provides a natural laboratory for studying the growth and feedback of PBHs embedded in dense, rotating media. Depending on how accretion and feedback unfold, such systems may evolve quasi-steadily, develop pronounced structural anomalies, or undergo energetic disruption, allowing stellar interiors to serve as unusually clean and stable environments for probing PBH physics.

The capture and subsequent evolution of PBHs within stars has been investigated across a wide range of stellar environments. An early application to primordial stars was given by \citet{Bambi2009}, who considered PBHs adiabatically contracted into Population III stars. In their scenario, PBHs sink by dynamical friction and form a central BH that can swallow the host. They concluded that PBHs with $\mbh\gtrsim 10^{22}\,{\rm g}$ could make the first stars short-lived. At the most compact end of the spectrum, PBH capture by neutron stars and white dwarfs has been studied extensively, as rapid PBH growth in these dense systems can lead to the destruction of the host star and thus potentially place constraints on the PBH abundance \citep{Kouvaris2014,Graham2015,Fuller2017,East2019,Genolini2020,Baumgarte2021,Baumgarte2024a,Baumgarte2024b,Richards2021,Schnauck2021,Caiozzo2024}.

At lower stellar densities, PBH capture by main-sequence stars leads to a qualitatively different evolutionary pathway. After capture, the PBH settles to the stellar center and begins accreting from the surrounding medium at a Bondi-like rate. This picture was explored early on by \citet{Markovic1995}, who treated the PBH as a point mass embedded in a star and examined how spherical accretion could modify stellar structure or lead to eventual destruction. \citet{Takhistov2019,Takhistov2022} proposed that PBH consumption of a rotating star can leave material outside the innermost stable circular orbit (ISCO), form an accretion disk, and power Blandford--Znajek jets \citep{Blandford1977}, producing GRB-, cocoon-, or microquasar-like transients. These works established the plausibility of PBH-powered disk outflows, but did not determine when an initially quasi-spherical PBH embedded in an ordinary star reaches the angular-momentum threshold for disk formation. More recently, the spherical accretion framework has been revisited using modern stellar evolution calculations. Employing self-consistent \texttt{MESA} models, \citet{Bellinger2023,Caplan2024} followed the long-term evolution of stars hosting central PBHs accreting quasi-spherically, demonstrating that feedback-regulated accretion can significantly alter stellar luminosities, lifetimes, and internal structure without necessarily leading to rapid disruption. In contrast, \citet{Baumgarte2026} considered the opposite limiting case in which accretion feedback and stellar rotation are neglected, allowing the PBH to grow rapidly at near-Bondi rates and ultimately consume the host star on short timescales.

A common limitation of these treatments is that the accretion efficiency of the embedded PBH is imposed rather than derived self-consistently from the properties of the accretion flow. For most of the evolution, quasi-spherical Bondi accretion is expected to be radiatively inefficient, while efficient energy extraction requires the formation of a rotationally supported accretion disk, which in turn depends on the angular-momentum content of the stellar gas. Because neither the angular-momentum supply nor the conditions for disk formation were explicitly modeled, the assumed efficiencies in these works span opposite limiting cases. A physically consistent description of PBH evolution within stars, therefore, requires tracking the different radiation regimes and angular momentum of the accreted gas to determine the regulated accretion and feedback on the star.  

In this paper, we develop a global framework for Hawking-star formation, evolution, and observables, following the system from PBH capture through in-stellar growth to either disk-driven disruption or quiet consumption. We begin by analyzing the possibility of two-body capture and rederive that single-pass capture via dynamical friction is inefficient for the PBH mass range of interest. Two-body dissipation nevertheless plays an essential role in the final stages of inspiral, once the PBH is already on a bound, star-crossing orbit. The initial seeding of bound orbits is instead mediated by three-body interactions: when the host star has a binary companion, such as a Jupiter analog, planetary slingshots can capture initially unbound PBHs, which subsequently migrate into the core through repeated dissipative transits. We derive a novel lower bound on the PBH mass for this combined capture-and-inspiral process to complete within a main-sequence lifetime.  

\begin{figure*}[]
  \centering
  \includegraphics[width=\textwidth]{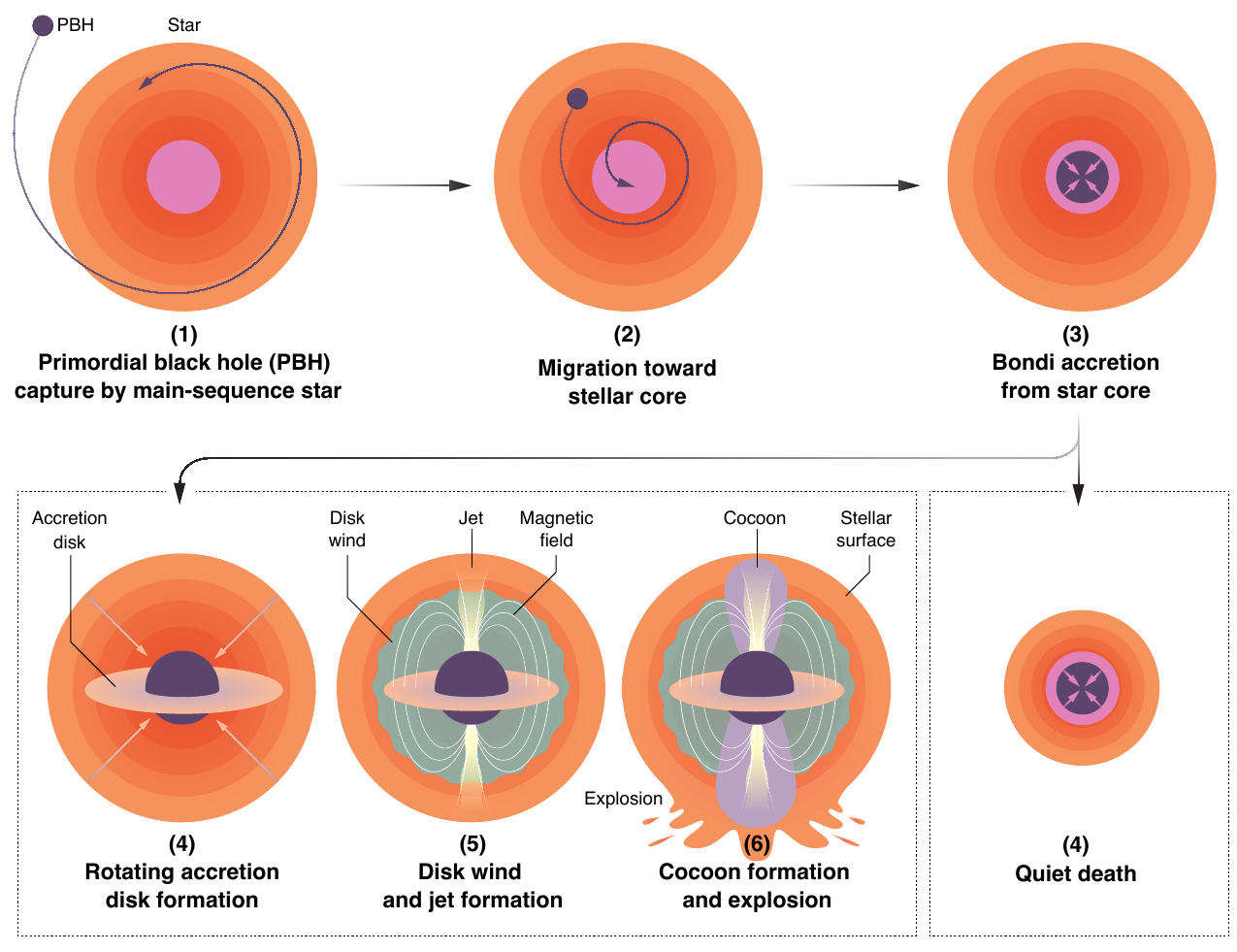}
  \caption{Schematic illustration (not to scale) of the evolutionary sequence of a Hawking star. An initially unbound PBH is first captured onto a bound, star-crossing orbit through three-body interactions, such as planetary slingshots in a stellar system with a binary companion, and subsequently migrates toward the stellar core through repeated dissipative transits (stages 1–2; see \S\ref{sec:capture} and \S\ref{sec:migration}). Once embedded in the core, the PBH grows through quasi-spherical Bondi accretion with low radiative efficiency (stage 3; see \S\ref{sec:bondi} and \citealt{Cantiello2026}). The subsequent fate depends on whether the PBH reaches the disk-formation threshold before consuming the host (see \S\ref{sec:fate}). If sufficient angular momentum is available when the PBH reaches the formal disk-formation mass, a rotationally supported accretion disk forms (explosive branch -- stage 4 on the left; see \S\ref{sec:disk}). Magnetic flux is subsequently generated in the disk, powering disk winds and PBH-driven jets (stage 5; see \S\ref{sec:feedback}), which inflate a hot cocoon and deposit energy into the stellar interior. This feedback ultimately disrupts the host star in an energetic explosion on timescales far shorter than its main-sequence lifetime (stage 6; see \S\ref{sec:explosion}). If, instead, the PBH is deposited after substantial stellar spin-down, or if its growth is sufficiently slow that disk formation is delayed until the stellar angular momentum has declined, the formal disk threshold may lie above the available stellar mass. In this quiet branch, the PBH consumes the star before an explosive disk forms, leaving a stellar-mass BH without a bright disk-powered transient.
  }
  \label{fig:sketch}
\end{figure*}

For the post-capture phase, we determine the accretion efficiency of PBHs embedded in stellar interiors self-consistently by explicitly tracking the angular-momentum content of the accreted gas. Using a combination of analytic arguments, detailed stellar evolution calculations, and 3D general-relativistic magnetohydrodynamic (GRMHD) simulations, we show that the evolution of such systems proceeds through well-defined stages that progressively alter the stellar interior. The first three stages in Figure~\ref{fig:sketch} illustrate the early evolutionary sequence. After a PBH is captured by a star and migrates into its core, accretion initially proceeds in a quasi-spherical Bondi regime with low radiative and mechanical efficiency prior to the formation of an accretion disk. We study this phase in depth in a companion paper, \citet{Cantiello2026}. As a result, feedback is strongly suppressed at early times, allowing the PBH to grow considerably more rapidly than in models that impose a sustained, non-negligible efficiency throughout the evolution \citep{Bellinger2023,Caplan2024}.

As the PBH mass increases, the Bondi radius grows and the specific angular momentum of the captured material inevitably rises. If the PBH reaches the angular-momentum threshold before consuming the host, a rotationally supported accretion disk forms once the circularization radius exceeds the ISCO. For rapidly rotating solar-type models, we find that this transition occurs when the PBH reaches a mass of $ \mbh \sim 10^{-2}-1\,\msun $ and dimensionless Kerr spin parameter of $a_\ast\approx 0.8$, independent of the initial PBH mass. More generally, however, the disk-formation mass depends on the host-star mass, the stellar rotation history, and the age at which the PBH is captured or deposited in the stellar interior. Disk formation marks a qualitative change in the system’s evolution: poloidal magnetic flux is generated in the disk, enabling efficient energy extraction and powering strong disk- and PBH-driven outflows. These outflows deposit energy into the stellar interior on timescales far shorter than the star’s main-sequence lifetime, leading to the rapid disruption of the host star. Conversely, PBHs captured too late, or with sufficiently small initial masses that they grow only after substantial stellar spin-down, may reach the formal disk-formation condition only after consuming most of the star, and therefore die quietly, leaving PBHs with masses comparable to that of the host. Thus, the fate of stars that capture PBHs bifurcates into explosive disruption or quiet consumption depending on the initial PBH mass, host mass, rotation history, and capture age.

This paper is intended as a first global framework for Hawking-star evolution rather than a definitive treatment of any single stage. The sequence from PBH capture to central growth, disk formation, explosive feedback, electromagnetic transients, and compact remnants spans several physical regimes that each warrant dedicated study. Our goal is therefore to identify the controlling physical transitions, derive the leading-order scalings, and demonstrate that Hawking stars need not evolve toward a single quiet endpoint. Instead, their fate bifurcates depending on whether the PBH forms an accretion disk before consuming the host.

The structure of this paper is as follows. We describe the gravitational capture of a PBH by a main-sequence star in \S\ref{sec:capture}, and three-body capture in \S\ref{sec:migration}. In \S\ref{sec:bondi}, we analyze the efficiency and early growth of the PBH through quasi-spherical Bondi accretion from the surrounding plasma. In \S\ref{sec:disk}, we identify the onset of efficient feedback, which occurs if and when the angular momentum of the accreted material exceeds that of the PBH’s ISCO and a compact accretion disk forms. In \S\ref{sec:fate}, we compare the quiet- and explosive-death populations and perform Monte Carlo population calculations to derive the corresponding PBH mass distributions. In \S\ref{sec:feedback}, we study the feedback in the disk-forming branch, as well as the resulting disk- and PBH-driven outflows. In \S\ref{sec:explosion}, we estimate the timescales and energetics of the stellar explosion powered by these outflows. In \S\ref{sec:numerical}, we combine \texttt{MESA} stellar evolution models with GRMHD simulations to follow the PBH’s evolution inside the stellar core and to validate our analytic predictions for Hawking stars that form disks. Finally, in \S\ref{sec:signatures} we discuss the observational implications and potential signatures of Hawking stars, and we summarize our conclusions in \S\ref{sec:conclusions}.

\section{Two-body Stellar Capture}\label{sec:capture} 

It has been suggested that the time-dependent gravitational potential during star formation can enhance PBH capture through adiabatic contraction \citep{Bambi2009,Capela_2013, Capela_2014, Esser2023}. However, these analyses assume an initial position-independent Maxwellian velocity distribution, rather than a phase-space distribution that depends on the gravitational potential of the protostellar cloud. As a result, a subset of PBHs is already on bound orbits \emph{before} the contraction takes place, so the time dependence mainly redistributes and compresses this pre-existing bound population rather than capturing PBHs from initially unbound trajectories. While we have not analyzed the star-formation scenario ourselves, it is not obvious to us that it should yield a significantly larger bound PBH population than the late-time capture discussed in the next section. In this section, we therefore focus on late-time capture by already-formed stars.

A PBH traversing a star loses orbital energy to the stellar medium via dynamical friction. If this energy loss were large enough, it could bind the PBH to the star on a single pass. However, this ``one-shot'' capture is ineffective for the PBH masses $\mbh$ of interest~\citep{Montero2019}. The energy lost in a \emph{single} transit is parametrically small, suppressed by the ratio $\mbh/M_\star$, where $M_\star$ is the stellar mass, and is insufficient to capture a typical halo PBH on a hyperbolic encounter. 

Dynamical friction instead plays an essential role in the post-capture inspiral process, once the PBH is already on a bound, star-crossing orbit. Repeated stellar transits then lead to cumulative energy loss, steadily shrinking the orbit until the PBH settles into the stellar core. In this sense, dissipation should be viewed as a mechanism that acts on already-bound orbits, rather than as an efficient capture mechanism --- we return to this in \S\ref{sec:migration}. Once the orbit is completely in the stellar interior, it sinks to the stellar core in a time significantly faster than typical (Bondi) accretion timescales~\citep{Montero2019}.  

One might worry that, for very light PBHs, convective stirring in fully convective hosts like late M dwarfs could prevent the PBH from remaining near the geometric center. Indeed, the instantaneous gravitational forcing from convective density fluctuations or large-scale flows may exceed the dissipative drag force itself. This does not necessarily prevent central settling, however, because once the PBH is fully embedded in the star the relevant competition is with the smooth, nearly harmonic restoring force of the stellar potential, rather than with the drag. Near the stellar center, the spherically averaged potential gives a restoring acceleration
\begin{equation}
    g \approx -\omega_0^2 r,
    \qquad
    \omega_0^2 = \frac{4\pi G\rho_0}{3}.
\end{equation}
A convective density fluctuation of size $\ell$ and amplitude $\delta\rho$ produces a stochastic gravitational acceleration
\begin{equation}
    a_{\rm stoch}\sim G\,\delta\rho\,\ell,
\end{equation}
corresponding to an equilibrium displacement
\begin{equation}
    r_{\rm wander}
    \sim
    \frac{a_{\rm stoch}}{\omega_0^2}
    \sim
    \frac{3}{4\pi}
    \left(\frac{\rho}{\rho_0}\right)
    \left(\frac{\delta\rho}{\rho}\right)\ell .
\end{equation}
For deep, subsonic convection, $\delta\rho/\rho\sim \mathcal{M}_c^2$, with $\mathcal{M}_c\equiv v_{\rm conv}/c_s\ll1$. Taking $\mathcal{M}_c\sim10^{-4}$ and $\ell\sim H_P\sim10^9$--$10^{10}\,\mathrm{cm}$ gives
\begin{equation}
    r_{\rm wander}\sim {\rm cm}\text{--}{\rm m},
\end{equation}
up to order-unity geometric factors. Such perturbations may drive stochastic wandering about the stellar center, but unless they correspond to order-unity density asymmetries on stellar-core scales, they are unlikely to eject the PBH from the central region. The main bottleneck therefore remains delivery onto a sufficiently tight, star-crossing orbit, not confinement once the PBH has reached the dense stellar interior.

For a PBH moving through a medium of density $\rho$ at velocity $\vect{v}$, the dynamical friction force is 
\citep{Chandr1949}
\begin{equation}
    \vect{F}_\text{df} = - 4\pi G^2 \mbh^2 \rho \ln \Lambda \frac{\hat{\vect{v}}}{v^2},
\end{equation}
where $\ln \Lambda$ is the Coulomb logarithm. The energy loss in a single stellar crossing is therefore
\begin{equation}
    \Delta E_\text{df} = \int_\text{transit} \dd \vect{s} \cdot \vect{F}_\text{df} = - 4\pi G^2 \mbh^2 \ln \Lambda \int_\text{transit} \dd s \, \frac{\rho(s)}{v(s)^2}.
\end{equation}
For an optimistic order-of-magnitude estimate, taking $v$ to be of order the stellar escape speed $v_{\text{esc},*} = \sqrt{2 G M_\star / R_\star}$ and the path length through the star to be of order its radius $R_\star$, one finds the parametric expression for the specific energy loss per transit:
\begin{equation}
\frac{\Delta E_\text{df}}{\mbh} \sim \frac{G \mbh}{R_\star} \ln \Lambda \sim v_{\text{esc},*}^2 \ln \Lambda \frac{\mbh}{M_\star}.
\end{equation}

An initially unbound PBH approaching the star with asymptotic speed $u$ has (positive) specific energy at infinity $\varepsilon_\infty = u^2/2$. One-pass capture by dynamical friction requires
\begin{equation}
    |\Delta E_{\text{df}}| \gtrsim \frac{1}{2}\mbh u^2,
\end{equation}
which implies a maximum capturable speed
\begin{equation}
    u 
    \lesssim u_\text{df}
    \equiv \left(\frac{2|\Delta E_\text{df}|}{\mbh}\right)^{1/2}
    \sim v_{\rm esc,\star}\left(2 \ln\Lambda\,\frac{\mbh}{M_\star}\right)^{1/2}.
\end{equation}
Because $\mbh \ll M_\star$, this $u_{\rm df}$ is extremely small: only an exceptionally tiny low-velocity tail of the halo distribution can be captured in a single pass. In a halo with characteristic dispersion $\sigma$, this ``slow'' fraction scales as
\begin{equation}
    f_{\rm slow}\sim \left(\frac{u_{\rm df}}{\sigma}\right)^3
\sim \left(\frac{v_{\rm esc,\star}}{\sigma}\right)^3
\left(2\ln\Lambda\,\frac{\mbh}{M_\star}\right)^{3/2},
\end{equation}
rendering one-shot capture extremely unlikely. 

Moreover, even in the optimistic near-parabolic limit $u\to 0$, a single transit would only yield a very weakly bound orbit. The post-transit semimajor axis is approximately
\begin{align}
a \approx \frac{G M_\star \mbh}{2|\Delta E_{\rm df}|}
\sim \frac{M_\star R_\star}{2 \mbh \ln\Lambda},
\end{align}
which can be enormous for light PBHs, implying orbital periods so long that subsequent star crossings, and thus cumulative dissipation, do not occur within a stellar lifetime. An external perturber would easily inhibit the inspiral by raising the periapsis out of the star or even unbinding the PBH from the star altogether \citep{Oncins2022}.

\section{Three-body Stellar Capture and Migration}\label{sec:migration}
Pure two-body gravity (PBH + star) cannot produce permanent capture: an initially unbound PBH conserves its energy at infinity, so it remains unbound unless there is dissipation or an additional degree of freedom. Since single-pass dissipation is negligible for the masses of interest, we focus on three-body capture mediated by a planet (or any other binary companion), followed by migration driven by dynamical friction during intermittent stellar transits.

At the planet's orbital radius $r\approx a_p$ (assumed to be constant, i.e.~circular), a PBH arriving from infinity with asymptotic heliocentric speed $u$ has local speed
\begin{equation}
v(a_p)
=\sqrt{u^2+\frac{2GM}{a_p}}
\approx \sqrt{u^2+2v_p^2},
\end{equation}
and can exchange energy with the planet through a gravitational slingshot.
The slingshot is controlled by the PBH's planet-frame hyperbolic-excess speed near $a_p$,
\begin{equation}
w_\infty^2=v^2+v_p^2-2vv_p\cos\alpha,
\end{equation}
where $\alpha$ is the angle between the PBH heliocentric velocity $\mathbf v$ and the planet's velocity $\mathbf v_p$ at encounter. In the planet frame, $|\mathbf w|$ is conserved but can be deflected by an angle $\theta$; the heliocentric specific-energy change is
\begin{equation}
\Delta\varepsilon \equiv \frac{u_f^2-u^2}{2}=\vect{v}_p\cdot(\vect{w}_f-\vect{w}_\infty),
\end{equation}
so the maximum possible magnitude occurs for a $180^\circ$ reversal ($\theta=\pi$),
\begin{equation}
|\Delta\varepsilon|_{\max}=2v_p w_\infty.
\end{equation}
Capture requires $\varepsilon_\infty=u^2/2 \lesssim |\Delta\varepsilon|_{\max}$, i.e.~
$u^2/2\lesssim 2v_p w_\infty$.
In the most favorable geometry, $\alpha=\pi$ so $w_\infty=v+v_p$ and the maximal energy loss is
$\Delta\varepsilon_{\min}=-2v_p(v+v_p)$.
This yields a hard upper bound on the asymptotic speed capturable in a single encounter,
\begin{equation}
u_\text{max}=2v_p\sqrt{3+\sqrt{10}}\approx 4.96\,v_p.
\end{equation}
We are effectively treating the planet as a point particle: for tight orbits, the maximum capturable speed in the planet's frame could instead be limited by the planet's escape velocity at its surface, but this will not parametrically change our arguments following Eq.~\eqref{eq:a_assumption}.

Away from this fine-tuned optimum, capture generically occurs for $u=\mathcal{O}(v_p)$ (up to order-unity factors from encounter geometry and deflection strength). Repeated encounters with the planet then drive chaotic diffusion of orbital elements and tend to equilibrate the accessible phase space on a characteristic timescale
\begin{equation}
t_{\rm capture}\sim P(a_p)\left(\frac{M_\star}{M_p}\right)^2,
\qquad
P(a)\equiv 2\pi\sqrt{\frac{a^3}{GM_\star}},
\end{equation}
with detailed balance between gravitational capture from and ejection into the DM halo, $t_{\rm capture} \sim 10^7\,{\rm yr}$ for the Sun-Jupiter system \citep{Peter2009,solar_basin_dynamics}.

We adopt the fully mixed approximation within this dynamically connected region, defined as the set of orbits with sufficient orbital energy, bounded by the inner edge:
\begin{equation}
a \gtrsim a_\text{min}= \frac{a_p}{2}, \label{eq:a_assumption}
\end{equation}
since highly eccentric orbits with aphelion near $a_p$ have $a\approx a_\text{min}$. In this mixed region, the distribution at fixed binding energy is approximately isotropic in angular momentum, and Liouville's theorem bounds the coarse-grained phase-space density by that of the ambient halo. We parameterize the local PBH mass density by $\rho_{\rm PBH}$ and take the halo velocity distribution to be Maxwellian with one-dimensional dispersion $\sigma$ (ignoring our boost through the halo for simplicity of argument), so that the mass-weighted velocity-space distribution function is
\begin{equation}
f(\mathbf v)
= \frac{\rho_{\rm PBH}}{(2\pi\sigma^2)^{3/2}}
\exp\!\left(-\frac{v^2}{2\sigma^2}\right),
\end{equation}
and $\int \dd^3 v\, f(\mathbf v)=\rho_\text{PBH}$, itself bounded from above by the DM mass density $\rho_\text{DM}$. \cite{Montero2019}  provided an upper bound on the rate of PBH deposition into the stellar interior, of 
\begin{align}
\dot{N}_\mathrm{max}
&\approx 16 \pi^2 G^2 M_\star \ln \Lambda [\max_v f] \label{eq:upper_bound} \\
&\approx \frac{1.5 \times 10^{-4}}{\mathrm{Gyr}}\,\frac{M_\star}{M_\odot}\,\frac{\rho_{\rm PBH} c^2}{1\,\mathrm{GeV\,cm^{-3}}}\,\left(\frac{\sigma}{10\,\mathrm{km\,s^{-1}}}\right)^{-3}\,\frac{\ln\Lambda}{30}. \nonumber
\label{eq:Ndotmax}
\end{align}
This is an exceptionally small rate except perhaps in the dense cores of ultra-faint dwarf galaxies with exceedingly small velocity dispersions~\citep{2025A&A...698A.290E}. In what follows, we will argue that even in the regime where the upper bound of Eq.~\eqref{eq:upper_bound} is comparable to the inverse lifetime of the stellar population, \emph{saturating} that upper bound is a tall order. Full stellar capture is only possible under the following conditions: (1) PBHs are relatively heavy with masses above some critical mass $M_{\rm BH}^{\rm crit}$; (2) the host star has tightly bound Jupiter analogs, or (3) there is a fluke orbital alignment between a nearby (perhaps minor) planetary companion and the PBH (if the first two conditions are not met). Therefore, full stellar capture of a PBH from the DM halo is a possible but exceptionally rare process.

Migration toward the stellar core proceeds because phase-space mixing repeatedly feeds PBHs into star-crossing orbits, where dynamical friction removes orbital energy. For an isotropic distribution at fixed $a$, the fraction of orbits with periapse $r_p<R_\star$ is
\begin{equation}
f_{\rm lc}(a)
\approx \left(\frac{j_{\rm lc}}{j_c}\right)^2
\approx \frac{2R_\star}{a},
\quad
j_{\rm lc}\approx \sqrt{2G M_\star R_\star},
\quad
j_c\approx \sqrt{G M_\star a},
\end{equation}
valid for $a\gg R_\star$. If each stellar transit removes energy $|\Delta E_{\rm df}|$, the orbit-averaged dissipation rate in the mixed regime is
\begin{equation}
\left\langle \frac{{\rm d}E}{{\rm d}t}\right\rangle
\sim -\,\frac{f_{\rm lc}(a)}{P(a)}\,|\Delta E_{\rm df}|.
\end{equation}
Using $E=-G M_\star \mbh/(2a)$, this implies a characteristic semimajor-axis decay time
    \begin{equation}
\label{eq:tau_df}
t_{\rm df}(a)\equiv \left|\frac{a}{\dot a}\right|
\sim \frac{1}{8\ln\Lambda}\,\frac{M_\star}{\mbh}\,P(a),
\qquad (a\gtrsim a_{\min}),
\end{equation}
up to order-unity factors from the stellar profile and the transit geometry. Because $P(a)\propto a^{3/2}$, the early evolution at the largest $a$ dominates the total inspiral time. Below $a_{\min}$, strong planetary resonances no longer fully mix the phase space, and secular perturbations may either drive the periapsis away from the star or cease to be star-crossing altogether; however, since $t_{\rm df}\propto a^{3/2}$ is a steeply rising function, the total inspiral time is dominated by the evolution near $a\approx a_{\min}$, and subsequent migration at tighter orbits is comparatively rapid. 

\begin{figure*}[t]
    \centering
    \includegraphics[width=\textwidth]{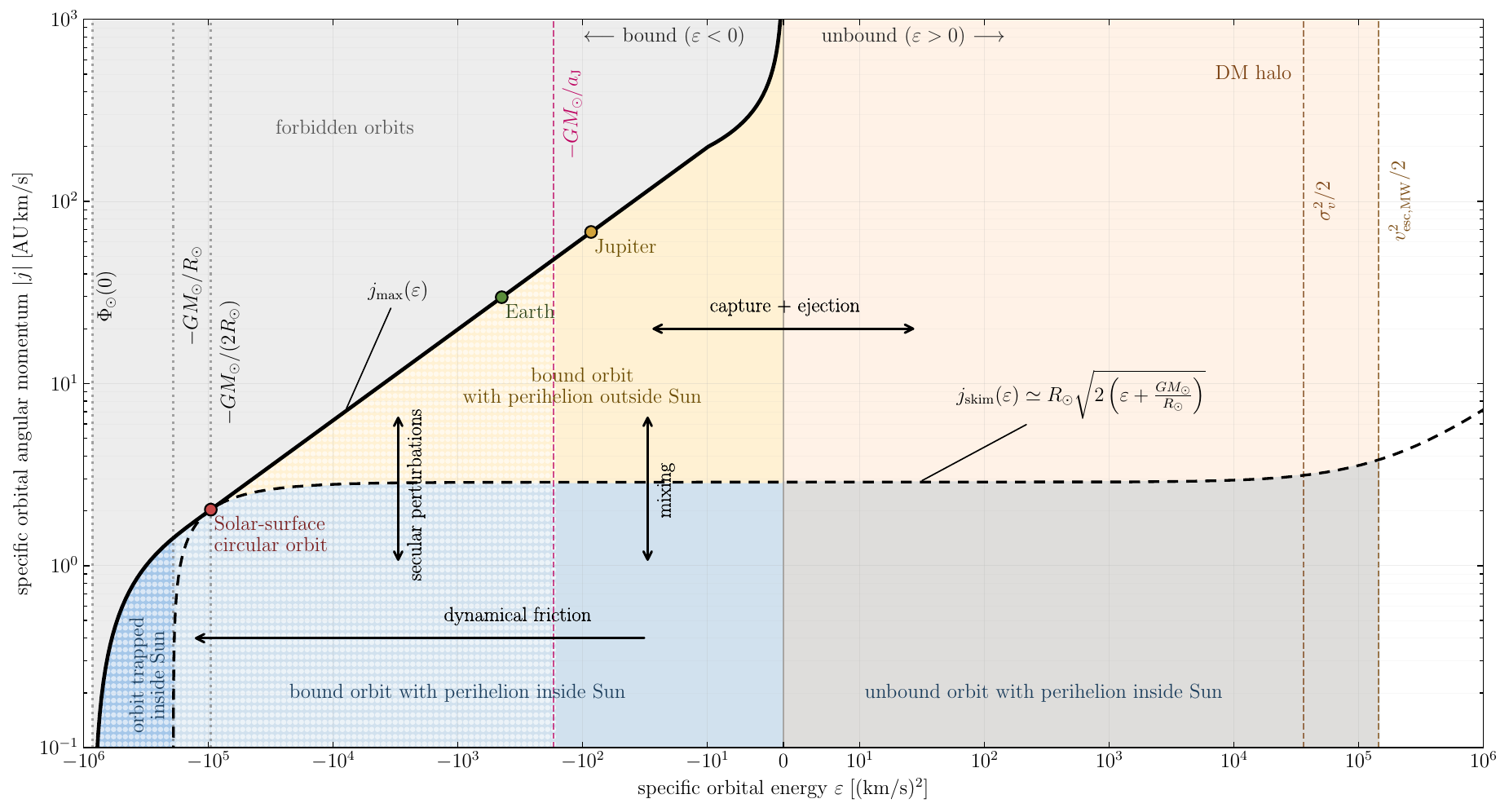}
    \caption{%
    Phase-space diagram of orbits in the Solar System (SS), in the plane of specific orbital energy $\varepsilon$ (symmetric log, linear for $|\varepsilon| < 10~\mathrm{km^2\,s^{-2}}$) and specific orbital angular momentum $|j|$ (log). The solid black curve is the maximum angular momentum $j_{\max}(\varepsilon)$ at fixed $\varepsilon$, attained by a circular orbit; orbits above are kinematically forbidden (gray). The dashed black curve is the locus $j_{\rm skim}(\varepsilon) \approx R_\odot \sqrt{2(\varepsilon + GM_\odot/R_\odot)}$ where the perihelion (aphelion) skims the Solar surface at $\varepsilon > -GM_\odot/R_\odot$ ($\varepsilon < -GM_\odot/R_\odot$). The light blue region below (tan region above) the dashed curve corresponds to orbits whose perihelion lies inside (outside) the Sun, while the dark blue region represents orbits entirely contained in the Sun. Bound phase space in the outer SS at $-GM_\odot/a_{\rm J} < \varepsilon < 0$ (right of vertical pink dashed line) is in detailed balance with the unbound ``bath'' of the DM halo (orange region) through Jovian slingshots (capture and ejection) of low-kinetic energy DM orbits much below the typical value of $\varepsilon \sim \sigma_v^2/2$ and maximum value $v_{\rm esc,MW}^2/2$. Angular momentum changes are rapid due to chaotic phase space mixing from secular perturbations and close planetary encounters. The phase space in the inner SS (left of pink line, polka dots) is sparsely populated by slow energy loss from dynamical friction, which occurs only for Sun-crossing orbits $|j| < j_\mathrm{skim}(\varepsilon)$, along with some diffusion due to gravitational scattering with minor planets. Eventually, dynamical friction would drain enough energy to deposit the PBH at the Solar core with $\varepsilon = \Phi_0(0)$ and $j = 0$. Reference orbits at Earth's and Jupiter's semi-major axes are marked.%
    }
    \label{fig:phase_space}
\end{figure*}

We depict the whole process described above in Figure~\ref{fig:phase_space}, for the prototypical example of the Solar System for specificity, although it should be noted that the PBH capture probability for the Sun is vanishingly small given the high velocity dispersion in the DM halo of the Milky Way, cf.~Eq.~\eqref{eq:upper_bound}.

Evaluating Eq.~\eqref{eq:tau_df} at $a=a_{\min}\approx a_p/2$ for a solar-type host with a Jupiter-analog companion ($a_p=5.2\,{\rm AU}$, $v_p=13\,{\rm km\,s}^{-1}$, $\ln\Lambda\approx 10$), we obtain

\begin{equation}
\begin{aligned}
t_{\rm df}(a_{\min})
&\approx \frac{P(a_{\min})}{8\ln\Lambda}\frac{M_\star}{\mbh} \\
&\approx 10^{10}\,{\rm yr}
\left(\frac{\mbh}{10^{22}\,{\rm g}}\right)^{-1}
\left(\frac{a_p}{5.2\,{\rm AU}}\right)^{3/2},
\end{aligned}
\label{eq:t_df}
\end{equation}
where we used $P(a_{\min})\approx(1/2)^{3/2}P_J\approx 4\,{\rm yr}$. The scaling with host-star and orbital parameters is $t_{\rm df}\sim M_\star P(a_{\min})\sim M_\star^{1/2}a_p^{3/2}$, so stars with wider planetary orbits or higher stellar masses require longer inspiral times for a PBH of fixed mass.

Setting $t_{\rm df}=t_\star\approx 10\,\text{Gyr}$ defines a critical PBH mass
\begin{equation}
M_{\rm BH}^{\rm crit} \sim 10^{22}\g \, \left(\frac{a_p}{5\,{\rm AU}}\right)^{3/2}
\left(\frac{M_\star}{M_\odot}\right)^{1/2}, \label{eq:m_bh_crit}
\end{equation}
above which inspiral completes within the main-sequence lifetime. For these parameters, this threshold falls at the upper end of the open asteroid-to-sublunar mass window $10^{17}$--$10^{23}\,{\rm g}$, where existing constraints on the PBH abundance are weakest. PBHs significantly more massive than $M_{\rm BH}^{\rm crit} $ can inspiral far more rapidly, making Hawking stars accessible even in young stellar populations. Light PBHs toward the lower end of the ``open'' mass window ($\sim 10^{17}\g$) have a sufficiently short decay time only for stars with planetary/binary companions at extremely tight orbits.

Not all stars have suitable companions on sufficiently tight orbits for the gravitational capture and subsequent timely inspiral of PBHs. The requirement $\mbh > M_{\rm BH}^{\rm crit}$ is for a \emph{typical} PBH; if it is not satisfied, the inspiral might still happen if the PBH is captured on a favorable trajectory. In summary, the capture and inspiral of a PBH is a quasi-random process---only a small fraction of stellar systems is expected to undergo this combined process, especially at masses significantly below the estimate of Eq.~\eqref{eq:m_bh_crit}.

The per-star Liouville upper bound \eqref{eq:upper_bound} can be combined with the companion fraction $f_\mathrm{comp}(\mbh)$ --- the probability that a randomly chosen star satisfies the joint conditions $t_{\rm capture} < t_\star$ and $t_{\rm df} < t_\star$ for a PBH of mass $\mbh$ --- to yield an order-of-magnitude cosmic capture rate
\begin{align}
    \Gamma_\mathrm{capture} \sim \bar{\rho}_\star^\mathrm{cosmic} \,\Big\langle \frac{f_{\rm comp}(\mbh) 
    \dot{N}_\star^{\rm max}}{M_\star} \Big\rangle \,, \label{eq:rate_capture}
\end{align}
with $\bar{\rho}_\star^\mathrm{cosmic} \approx 5 \times 10^8 \, M_\odot \, \mathrm{Mpc}^{-3}$ the cosmic stellar mass density \citep{Madau2014} and the average taken over all stellar systems in all environments. Evaluated at solar-neighborhood values ($\rho_{\rm DM} = 0.4\,\mathrm{GeV\,cm^{-3}}$, $\sigma = 200\,\mathrm{km\,s^{-1}}$, with $\ln\Lambda = 30$, $f_{\rm PBH}=1$), the bound is $\dot{N}_\star^{\rm max}/M_\star \approx 7\times10^{-18}\,\mathrm{yr^{-1}}\,M_\odot^{-1}$.

The companion fraction is controlled by $a_{p,\rm max}(\mbh) \approx 5\,{\rm AU}\,(\mbh/10^{22}\,{\rm g})^{2/3}$, the maximum companion semimajor axis for which dynamical-friction inspiral completes within a stellar lifetime. For $\mbh = 10^{-11}\,M_\odot$, $a_{p,\rm max}\sim 8\,$AU admits both the giant-planet \citep{Cumming2008} and the close-stellar-binary \citep{Moe2017} populations, yielding $f_{\rm comp}\sim 0.05$--$0.3$ depending on whether only Jupiter analogs are counted or all close companions are admitted. For $\mbh = 10^{-14}\,M_\odot$, $a_{p,\rm max}\sim 0.08\,$AU restricts qualifying systems to hot Jupiters/Neptunes, super-Earths, and short-period stellar binaries, for which we estimate $f_{\rm comp}\sim 10^{-2}$--$0.2$ \citep{Mayor2011, Howard2012}. For $\mbh = 10^{-16}\,M_\odot$, $a_{p,\rm max}\sim 0.004\,{\rm AU}\,(M_\star/M_\odot)^{1/2}$ lies interior to a solar-type star but corresponds to a few stellar radii for late M dwarfs, admitting only the most compact ultra-short-period planetary systems \citep{SanchisOjeda2014} and giving $f_{\rm comp}\sim 10^{-3}$--$10^{-2}$.

Two further systematic uncertainties bracket Eq.~\eqref{eq:rate_capture} over several orders of magnitude. First, while luminous galaxies with Milky-Way-like $\rho_{\rm DM}/\sigma^3$ host the bulk of the cosmic stellar mass, dwarf galaxies enhance the per-star Liouville bound by factors $(\rho_{\rm DM}/\sigma^3)/(\rho_{\rm DM}/\sigma^3)_\mathrm{MW}\sim 10^{4}\text{--}10^{5}$. Even though dwarfs contribute only a few percent of $\bar{\rho}_\star$, this can shift $\Gamma_\mathrm{capture}$ upward by $\mathcal{O}(10\text{--}100)$. Second, saturation of Eq.~\eqref{eq:upper_bound} requires complete chaotic mixing of the bound-orbit phase space, which is unlikely to be exact and may suppress the rate by $\mathcal{O}(0.1)$. Combining these effects, our pessimistic and optimistic estimates of $f_\mathrm{comp}$ and $\Gamma_\mathrm{capture}$ differ by $\sim 3$--$4$ orders of magnitude, as listed in Tab.~\ref{tab:rate_table}.
\begin{table}[h]
\centering
\renewcommand{\arraystretch}{1.2}
\begin{tabular}{c|cc|cc}
\hline
$\mbh\,[M_\odot]$ & $f_{\rm comp}^{\rm pess}$ & $f_{\rm comp}^{\rm opt}$ &
$\Gamma_\mathrm{capture}^\mathrm{pess}$ & $\Gamma_\mathrm{capture}^\mathrm{opt}$ \\
\hline
$10^{-16}$ & $10^{-3}$ & $10^{-2}$ & ${\sim}\,3\!\times\!10^{-13}$ & ${\sim}\,3\!\times\!10^{-9}$ \\
$10^{-14}$ & $10^{-2}$ & $0.2$     & ${\sim}\,3\!\times\!10^{-12}$ & ${\sim}\,7\!\times\!10^{-8}$ \\
$10^{-11}$ & $0.05$    & $0.3$     & ${\sim}\,2\!\times\!10^{-11}$ & ${\sim}\,10^{-7}$ \\
\hline
\end{tabular}
\caption{Pessimistic and optimistic estimates of the companion fraction, $f_{\rm comp}$, and capture rate, $\Gamma_\mathrm{capture}$, for different PBH masses, $\mbh$. The capture rate is in units of $\mathrm{yr}^{-1}\,\mathrm{Mpc}^{-3}$ and scales linearly with $f_{\rm PBH}$, normalized here to unity.}
\label{tab:rate_table}
\end{table}
For $\mbh = 10^{-16}\,M_\odot$, the Bondi growth time at solar-core conditions $t_{\rm B} \approx 10^{10}\,\mathrm{yr}\,(\mbh/10^{-16}\,M_\odot)^{-1}$ is comparable to a Hubble time, so explosion events lag captures by a similar interval and the present-day explosion rate is set by the cosmic-history-averaged $\Gamma_\mathrm{capture}$ rather than its instantaneous value.

\section{Accretion}\label{sec:bondi}

When a PBH settles at the center of a star, it begins to grow through quasi-spherical accretion of stellar material. For PBHs in the asteroid mass range, the Bondi radius
\begin{equation}\label{eq:Rb}
    \rb = \frac{2 G \mbh}{c_s^2} \approx 4 \times 10^{-3}\,\cm\,\left(\frac{\mbh}{10^{20}\,\text{g}}\right) \left(\frac{c_s}{600\,{\rm km\,s^{-1}}}\right)^{-2}
\end{equation}
is far smaller than the stellar circularization radius, so no accretion disk forms, and the flow remains spherical. This Bondi phase does not require gas to fall ballistically from stellar-core scales directly onto the event horizon. It only requires that gas captured at $\rb$ have insufficient specific angular momentum to circularize outside the ISCO. The angular momentum of gas at $\rb$ is insufficient to circularize outside the ISCO, and in the absence of viscosity or angular-momentum transport, the Bernoulli parameter remains negative throughout. No outflows or jets are launched, and infalling material advects its energy into the horizon. The magnetic field is dynamically negligible throughout this phase: both the instantaneous field strength ($p_{\rm mag}/p_{\rm ram} \lesssim 10^{-11}$) and any flux accumulated via reconnection-limited buildup ($\beta_{p,\rm eq} =p_{\rm gas}/p_{\rm mag}\sim 10^8$) fall orders of magnitude short of dynamical importance \citep[see][]{Cantiello2026}, where $p_{\rm mag,ram,gas}$ are the magnetic, ram, and gas pressures of the flow.

We treat the detailed physics of the accretion flow, including relativistic bremsstrahlung and pair processes, Klein--Nishina corrections, radiative feedback on the Bondi boundary conditions, and the transition to hyper-critical regime, in our companion paper, \citet{Cantiello2026}. The key results of that analysis, which we adopt here, are as follows. The accretion flow passes through three physically distinct regimes as $\mbh$ increases: (1) a collisionless regime\footnote{For the lightest PBHs, the inner flow may become collisionless well inside $\rb$, but the gas remains collisional at the Bondi scale where $\dot{M}_{\rm BH}$ is determined; Appendix~B of \citet{Cantiello2026} shows that the resulting loss-cone/recycling correction reduces the Bondi rate by at most tens of percent.} ($\mbh \lesssim 10^{-14}\,M_\odot$) where the Bondi sphere is optically thin and the radiative efficiency $\eta \sim 0.1$--$0.01$; (2) a bremsstrahlung cooling regime ($10^{-14}$--$10^{-12}\,M_\odot$) where cooling drives the flow toward isothermal and $\eta$ reaches a minimum $\approx 10^{-2}$; and (3) a hyper-critical, photon-trapped regime ($\mbh \gtrsim 10^{-12}\,M_\odot$) where accretion proceeds at the full Bondi rate.

The efficiency $\eta \sim 10^{-2}$ is roughly an order of magnitude below the thin-disk value $\eta = 0.08$ adopted in earlier work \citep{Bambi2009,Bellinger2023}. This low efficiency has two important consequences for the present paper. First, radiative and convective feedback remain too weak to suppress accretion below the Bondi rate. Second, once the Bondi sphere becomes optically thick, photon trapping limits the escaping luminosity but does not impose an Eddington cap on the mass accretion rate in spherical geometry. The PBH therefore continues to grow approximately at the Bondi rate throughout the spherical phase until disk formation or stellar consumption. The growth bottleneck is instead set by the slow early Bondi phase at the smallest masses, yielding a critical initial PBH mass $M_0\sim10^{-16}\,M_\odot$ for a solar-type star to reach a main-sequence endpoint within a Hubble time, $t_H$. This threshold is well below the mass scales emphasized in previous studies that adopted more luminous spherical-accretion feedback prescriptions \citep{Bambi2009,Bellinger2023}.

\section{Disk formation}\label{sec:disk}

A rotationally supported disk forms only once the specific angular momentum of gas captured within the Bondi radius exceeds the specific angular momentum required to circularize outside the PBH’s ISCO. In principle, turbulent motions could transiently impart sufficient angular momentum to form a disk and generate nonzero feedback. Convective motions should therefore be evaluated on the scale that is actually captured by the PBH. A bulk convective velocity on scales much larger than $\rb$ mainly acts as a relative velocity between the gas and the PBH, rather than as rotation of the gas about the PBH. The angular momentum of the accreted material is instead set by the velocity difference across the Bondi sphere, i.e., by vorticity or turbulent eddies on scales $\sim\rb$. Because $\rb$ is much smaller than the convective outer scale $L_{\rm conv}$, this differential velocity, and therefore the associated specific angular momentum, is strongly suppressed compared with a naive extrapolation from stellar-core scales. For an isotropic Kolmogorov-turbulent velocity field, this gives only random, small-scale angular momentum,
\begin{equation}
    j \sim r\,v_{\rm turb}(r) \sim r^{4/3},
\end{equation}
which corresponds to a circularization radius
\begin{equation}
    R_{\rm circ} \sim \frac{j^2}{G\mbh},
\end{equation}
that lies far inside the PBH's ISCO. Thus, turbulent fluctuations alone cannot provide the coherent, large-scale angular momentum needed to form a centrifugally supported disk. We confirm this conclusion with a 3D GRMHD simulation that employs Maxwellian-distributed turbulent velocity fields initialized around the PBH.

As the PBH grows, two competing trends govern the onset of disk formation. On the one hand, the Bondi radius increases, allowing the PBH to capture material with progressively higher specific angular momentum. On the other hand, the ISCO radius also grows with PBH mass, increasing the threshold angular momentum required for circularization. In the local Bondi-fed problem considered here, the former effect wins: $j_B/j_{\rm ISCO}\propto \mbh$, so disk formation occurs above a critical PBH mass.\footnote{The opposite scaling in the $M_+$ estimate of \citet{Markovic1995} follows from a torque-free Lagrangian-shell assumption: gas initially at $r_0$ retains $j_0=\Omega r_0^2$ throughout its inward drift, and its arrival is associated with $M_{\rm BH}=M_{\rm enc,0}(r_0)\propto r_0^3$, giving $j_0/j_{\rm ISCO}\propto M_{\rm BH}^{-1/3}$. This retained-shell prescription overestimates the coherent angular momentum delivered in the quasi-hydrostatic stellar core considered here. Our \texttt{MESA} models include magnetic and rotational angular-momentum transport that maintains near-solid-body rotation, so gas supplied inward exchanges angular momentum with the surrounding star and reaches the Bondi region with the local corotation value $j_B\simeq\Omega \rb^2$.} This marks the transition from quasi-spherical Bondi accretion to rotationally supported, disk-mediated accretion.
We adopt the working criterion
\begin{equation}\label{eq:j}
    j_{\rm B}(t) \;\gtrsim\; j_{\rm ISCO}(t),
\end{equation}
where $j_{\rm ISCO}(t)=\tilde{L}_{\rm ISCO}(a_\ast(t))G\mbh(t)/c$, and the dimensionless ISCO angular momentum, $\tilde{L}_{\rm ISCO}(a_\ast)$, varies from $\tilde{L}_{\rm ISCO,min} = 2/\sqrt{3}$ for a maximally spinning PBH to $\tilde{L}_{\rm ISCO,max} = 2\sqrt{3}$ for a Schwarzschild PBH \citep{Bardeen1970}. The quantity $j_{\rm B}$ is the mean specific angular momentum of stellar gas at the Bondi radius. For a uniformly rotating spherical shell, the shell-averaged value is
\begin{equation}
    j_{\rm B}(t) \approx \frac{2}{3}\,\Omega_\ast(t)\,\rb(t)^2
    = \frac{8}{3}\,\frac{\Omega_\ast(t)\,G^2\,\mbh(t)^2}{c_s^4(t)}\,,
    \label{eq:jB}
\end{equation}
where $\Omega_\ast$ is the local stellar angular velocity.

Magnetic torques and rotational instabilities drive the stellar interior toward rigid rotation throughout the Bondi growth phase (see \S\ref{sec:numerical}), so the angular velocity at the Bondi radius equals the bulk stellar rotation rate. Parameterizing this by the dimensionless surface rate $\lambda(t) \equiv \Omega_\ast(t)/\Omega_{\rm k} \le 1$ relative to the breakup frequency $\Omega_{\rm k}\equiv \sqrt{GM_\star/R_\star^3}=\sqrt{4\pi G\bar\rho_\star/3}$, where $\bar\rho_\star \approx 1.4\,{\rm g\,cm^{-3}}$ is the \emph{mean} stellar density for a Sun-like star (distinct from the central density $\rho_0 \approx 150\,{\rm g\,cm^{-3}}$ that sets the local Bondi rate),
\begin{equation}\label{eq:Omega}
    \Omega_\ast(t) = \lambda(t)\,\sqrt{\frac{4\pi G\bar\rho_\star}{3}}\,.
\end{equation}

The first shell to circularize outside the ISCO is obtained when $j_{\rm B} = j_{\rm ISCO}$. Substituting Eqs.~\eqref{eq:Rb} and \eqref{eq:Omega} into this condition yields the PBH mass at the time of disk formation, $t=t_d$,
\begin{equation}
    M_{\rm BH,disk} = \frac{3}{8}\,
    \frac{\tilde{L}_{\rm ISCO}(a_\ast(t_d))c_s^4(t_d)}{\Omega_\ast(t_d) Gc}=
    \frac{9~(3)}{8\sqrt{\pi\bar\rho_\star}}\,
    \frac{c_s^4(t_d)}{\lambda(t_d)\,c\,G^{3/2}}\,,
    \label{eq:Mcrit}
\end{equation}
where the coefficient $9\,(3)$ corresponds to a non-spinning (maximally spinning) PBH.

To derive the PBH spin upon disk formation, we recall that before disk formation, the accreted gas falls directly onto the PBH, carrying its angular momentum $j_{\rm B}$. Since the initial PBH mass is negligible compared to $M_{\rm BH,disk}$, we use Eq.~\eqref{eq:jB} to estimate the accumulated BH angular momentum at disk formation as
\begin{equation}
    J_{\rm BH,disk} = \int_0^{M_{\rm BH,disk}} j_{\rm B}(M)\,\dd M
    \approx \frac{8}{9}\,\frac{\Omega_\ast(t_d)\,G^2\,M_{\rm BH,disk}^3}{c_s^4(t_d)}.
\end{equation}
The corresponding PBH spin parameter is therefore
\begin{equation}\label{eq:a_from_J}
    a_{\ast,{\rm disk}} = \frac{c\,J_{\rm BH,disk}}{G\,M_{\rm BH,disk}^2}
    = \frac{8}{9}\,\frac{c\,\Omega_\ast(t_d)\,G\,M_{\rm BH,disk}}{c_s^4(t_d)}\,.
\end{equation}

Substituting the instantaneous disk-formation condition, Eq.~\eqref{eq:Mcrit}, into Eq.~\eqref{eq:a_from_J}, all dimensional quantities cancel,
\begin{equation}\label{eq:a}
    a_\ast = \frac{8}{9}\,\frac{c\,\Omega_\ast(t_d)\,G}{c_s^4(t_d)}
    \;\cdot\;
    \frac{3}{8}\,\frac{\tilde{L}_{\rm ISCO}(a_\ast(t_d))\,c_s^4(t_d)}{\Omega_\ast(t_d)\,G\,c}
    = \frac{\tilde{L}_{\rm ISCO}(a_\ast(t_d))}{3}\,.
\end{equation}
Thus for $M_0\ll M_{\rm BH,disk}$, the spin at disk formation is determined by a fixed-point equation that depends solely on the Kerr metric. For prograde accretion onto a Kerr BH, this equation has a unique solution $a_\ast \approx 0.8$. The cancellation arises because a larger $\Omega_\ast$ leads to disk formation at a lower PBH mass, while the accumulated spin scales in such a way that these dependencies compensate.

Using $a_\ast \approx 0.8$ in Eq.~\eqref{eq:Mcrit}, the
instantaneous PBH mass required for disk formation is
\begin{equation}
    \begin{split}
    M_{\rm BH,disk} &\approx 8\times 10^{-2}\,\msun
    \left(\frac{0.1}{\lambda(t_d)}\right)\times\\
    &
    \left(\frac{c_s(t_d)}{6.9\times10^{7}\,{\rm cm\,s^{-1}}}\right)^{4}
    \left(\frac{\bar\rho_\star}{1.4\,{\rm g\,cm^{-3}}}\right)^{-1/2}.
    \end{split}
\end{equation}
Thus, while all dimensional quantities cancel in the spin estimate, yielding the nearly universal solution $a_\ast\approx0.8$, the PBH mass at disk formation depends on the stellar evolution track through $\lambda(t_d)$ and $c_s(t_d)$. The latter varies comparatively modestly during the relevant evolution, whereas stellar spin-down can change $\lambda$ by orders of magnitude. For PBHs captured in young, rapidly rotating solar-type stars, this gives $M_{\rm BH,disk}\sim10^{-2}$--$10^{-1}\,\msun$. However, stellar spin-down can substantially increase $M_{\rm BH,disk}$ for PBHs captured late, or for low-mass seeds with long growth lags. This distinction is important for determining the fate of the system. If the disk-formation mass exceeds the available stellar mass, $M_{\rm BH,disk} \gtrsim M_\star$, then Eq.~\eqref{eq:j} is not satisfied before the PBH consumes the host. In this case, no explosive accretion disk forms and the star dies quietly. We characterize the resulting disk-forming and quiet-death populations in \S\ref{sec:fate}.

The above derivation of the PBH mass and spin at disk formation captures the disk-formation threshold under the assumptions most relevant to the early, low-angular-momentum phase: (i) angular momentum is effectively conserved along streamlines because the inflow time across $\rb$ is short compared to viscous or magnetic transport timescales; (ii) the stellar core rotates nearly as a solid body; and (iii) the local flow is characterized by $(\rho_0,\,c_s)$. For $\mbh\!<\!M_{\rm BH,disk}$, the accretion proceeds quasi-spherically and is radiatively inefficient (Bondi-like). As the PBH grows beyond this threshold, disk formation enables viscous dissipation, and the radiative efficiency increases dramatically.

\section{Quiet and explosive fates}\label{sec:fate}
Disk formation requires $M_{\rm BH,disk}\lesssim M_\star$. Thus, from Eq.~\eqref{eq:Mcrit}, the condition for disk formation is
\begin{equation}
    \lambda(M_\star,t_{\rm crit})>0.9{c_s^4(M_\star,t_{\rm crit})
    \over
    cG\, \Omega_\mathrm{k}(M_\star,t_{\rm crit})M_\star},
    \label{eq:lambda_crit}
\end{equation}
where stellar spin-down makes $\lambda$ decrease monotonically with age, and $t_{\rm crit}$ is the critical stellar age at which the spin has declined to the threshold value. Thus, a PBH deposited in the stellar interior at age $t_{\rm cap}$ will form a disk only if
\begin{equation}
    t_{\rm cap}+t_{\rm lag}(M_0,M_\star) < t_{\rm crit} (M_\star),
    \label{eq:disk_success}
\end{equation}
where $t_{\rm lag}$ is the time between PBH deposition and disk formation. For sufficiently massive seeds, it can be short compared with the main-sequence lifetime, while for $M_0\sim10^{-16}M_\odot$ it can be comparable to a Hubble time.

We now compute the capture-weighted fraction of Hawking stars that satisfy Equation~\eqref{eq:disk_success}. Let $\xi(M_\star)= \dd N/\dd M_\star$ be the stellar initial mass function and
\begin{equation}
    t_\star(M_\star)\equiv
    \min\left[t_{\rm MS}(M_\star),t_H\right],
\end{equation}
the time over which the star is available for capture. For a time-independent capture, the capture time at fixed stellar mass is uniformly distributed over the interval $0<t_{\rm cap}<t_\star$. The successful capture-time interval is therefore
\begin{equation}
    {\cal T}_{\rm disk} = \label{eq:Tdisk_def} \int_0^{t_\star} \dd t_{\rm cap}\,\Theta\!\left[\min\left(t_\star,t_{\rm crit}\right)-t_{\rm lag}(M_0)-t_{\rm cap}\right].
\end{equation}
Since the integrand is unity only for $0<t_{\rm cap}<t_{\rm crit}-t_{\rm lag}$, the integral simply returns the length of the intersection of this interval with $0<t_{\rm cap}<t_\star$:
\begin{equation}
    {\cal T}_{\rm disk}(M_\star)=\max\left[\min\left(t_\star(M_\star),\,t_{\rm crit}(M_\star)\right)-t_{\rm lag}(M_0,M_\star),0\right].
    \label{eq:Tdisk_simple}
\end{equation}

Since Eq.~\eqref{eq:upper_bound} gives $\dot N_{\max}\propto M_\star$, while the number of stars in a mass bin is proportional to $\xi(M_\star)$, the capture rate in that bin is weighted by stellar mass density,
\begin{equation}
    \dd \Gamma_\mathrm{capture}
    \propto
    M_\star\,\xi(M_\star)\,
    f_{\rm comp}(M_\star,M_0)\, \dd M_\star .
    \label{eq:capture_mass_weighting}
\end{equation}
where $f_{\rm comp}$ is the fraction of stars with a companion architecture capable of capturing and delivering a PBH of initial mass $M_0$ to the stellar interior within $t_\star$.

The population-averaged disk-forming fraction is therefore
\begin{equation}
    F_{\rm disk}(M_0)={\displaystyle
    \int \dd M_\star\,M_\star \xi(M_\star)\,f_{\rm comp}(M_\star,M_0)\,{\cal T}_{\rm disk}(M_\star)
    \over
    \displaystyle\int \dd M_\star\,M_\star \xi(M_\star)\,f_{\rm comp}(M_\star,M_0)\,t_\star(M_\star)
    }.
    \label{eq:population}
\end{equation}

\subsection{Monte Carlo implementation}
We evaluate Eq.~\eqref{eq:population} with a Monte Carlo population calculation. For each initial PBH mass $M_0$, we draw host stars from the capture-weighted distribution
\begin{equation}
    p(M_\star|M_0)={1\over {\cal N}_{\rm cap}(M_0)}{\dd \Gamma_\mathrm{capture}\over \dd M_\star}t_\star(M_\star),
    \label{eq:pMstar_MC}
\end{equation}
where
\begin{equation}
    {\cal N}_{\rm cap}(M_0)=\int dM_\star\,{\dd \Gamma_\mathrm{capture}\over \dd M_\star}t_\star(M_\star)
\end{equation}
normalizes the distribution. The factor of $t_\star$ accounts for the time interval over which captures can occur. At fixed $M_\star$, we draw the capture time uniformly over $0<t_{\rm cap}<t_\star(M_\star)$, corresponding to a time-independent capture rate. We model stellar spin-down using a phenomenological fit to the median rotation period tracks across four low-mass stellar bins. The data used in the fit are the median cluster rotation periods as a function of age \citep{Rebull2016,Rebull2017,Rebull2018}, using the young USco/$\rho$ Oph population, the Pleiades, Praesepe, and, for the solar-type bin, the present Sun. For each mass bin, we approximate the post-turnover spin-down branch by
\begin{equation}
    \label{eq:lambda}
    \lambda(M_\star,t)=\lambda_0(M_\star)\left[1+f_{\rm br}(M_\star){t\over t_{\rm br}(M_\star)}\right]^{-\beta(M_\star)} .
\end{equation}
The fitted parameters used in the fiducial calculation are\footnote{These fits are intended to capture the median post-turnover evolution, not the full observed rotation distribution or the early spin-up phase of low-mass stars.} listed in Tab.~\ref{tab:lambda}.

\begin{table}[h]
\centering
\begin{tabular}{c|cccc}
$M_\star/M_\odot$ & $0.08\!-\!0.20$ & $0.20\!-\!0.35$ & $0.35\!-\!0.60$ & $\geq0.60$ \\
\hline
$\lambda_0$ & 0.49 & 0.12 & 0.055 & 0.058 \\
$t_{\rm br}\ [{\rm Gyr}]$ & 0.30 & 0.12 & 0.050 & 0.008 \\
$\beta$ & 1.30 & 0.90 & 0.70 & 0.55 \\
\end{tabular}
\caption{Parameters for Eq.~\eqref{eq:lambda} across stellar mass bins: $\lambda_0$ the Keplerian rotation fraction before the braking time, $t_\mathrm{br}$ the characteristic braking time, and $\beta$ the power-law index governing the spin-down rate.}
\label{tab:lambda}
\end{table}

To suppress magnetic braking above the Kraft break \citep{Kraft1967,Beyer2024}, we multiply the spin-down term by
\begin{equation}
f_{\rm br}(M_\star)=
\begin{cases}
1, & M_\star\leq1.1\,\msun,\\
\frac{1.4\,\msun-M_\star}{0.3\,\msun}, &
1.1\,\msun<M_\star<1.4\,\msun,\\
0, & M_\star\geq1.4\,\msun,
\end{cases}
\end{equation}
so cool stars experience full braking, stars above $1.4\,\msun$ retain approximately their initial rotation, and the transition is interpolated between these limits.

We use Eq.~\eqref{eq:lambda} as a deterministic median spin-down prescription. For each stellar-mass bin, $\lambda_0$, $t_{\rm br}$, and $\beta$ are fixed to the values in Table~\ref{tab:lambda}. These parameters are intended to reproduce representative post-turnover rotation evolution, calibrated to median cluster rotation periods and the present solar rotation, rather than the full observed spread of young-star rotation rates. The braking time should therefore be interpreted as an effective calibration parameter for the adopted median track, not as a physical wind-braking timescale.

For each realization, the PBH reaches the formal disk-formation threshold at
\begin{equation}\label{eq:t_d}
    t_d=t_{\rm cap}+t_{\rm lag}(M_0,M_\star),
\end{equation}
provided this time lies within the modeled stellar lifetime. We then evaluate the disk-formation mass along the corresponding stellar evolution and rotation track. The system is classified as an explosive disk-forming event if
\begin{equation}
    t_d<t_\star(M_\star),
    \qquad
    M_{\rm BH,disk}<M_\star .
\end{equation}
In this case, the final PBH mass is taken to be $M_f\approx M_{\rm BH,disk}$ because the subsequent disruption occurs on a timescale much shorter than the PBH mass-doubling time.

If the $M_{\rm BH,disk}>M_\star$, the PBH would consume the host before forming an explosive disk. We therefore classify the system as a quiet death if the PBH reaches the stellar mass before the end of the modeled stellar lifetime. In this branch, the terminal PBH mass is of order the consumed stellar mass, $M_f\approx M_\star$. If the PBH reaches neither disk formation nor quiet consumption before $t_\star$, we classify the system as having no main-sequence endpoint in the present calculation.

The PBH mass distribution is constructed directly from the Monte Carlo realizations. Since host masses are sampled directly from the capture-weighted distribution in Eq.~\eqref{eq:pMstar_MC}, and capture times are drawn uniformly over $0<t_{\rm cap}<t_\star(M_\star)$, all Monte Carlo realizations have equal statistical weight. The resulting histogram estimates
\begin{equation}
    {\dd F_{\rm exp}\over \dd \log M_f}(M_0)\approx{1\over N_{\rm tot}\,\Delta\log M}\sum_{i\in{\rm exp}}{\cal I}_i(M_f),
    \label{eq:mc_mass_distribution}
\end{equation}
where $N_{\rm tot}$ is the total number of Monte Carlo realizations for the fixed initial PBH mass $M_0$, $\Delta\ln M$ is the logarithmic bin width, and ${\cal I}_i(M_f)=1$ if the final mass $M_{f,i}$ of realization $i$ lies in the bin centered on $M_f$, and zero otherwise. With this normalization, the area under the histogram is the explosive fraction $F_{\rm exp}(M_0)$, not unity. We repeat the same procedure separately for each value of $M_0$.

We adopt a Kroupa initial mass function \citep{Kroupa2001} over the full stellar mass range sampled in the
Monte Carlo,
\begin{equation}
    \xi(M_\star)\equiv {\dd N\over \dd M_\star}
    \propto
    \begin{cases}
    M_\star^{-1.3}, & 0.08\,\msun < M_\star < 0.5\,\msun,\\
    M_\star^{-2.3}, & 0.5\,\msun < M_\star.
    \end{cases}
\end{equation}
with the normalization chosen so that $\xi(M_\star)$ is continuous at $0.5\,\msun$. Since we compute capture-weighted fractions at fixed $M_0$, the overall normalization of the IMF cancels between numerator and denominator in Eq.~\eqref{eq:population}.

To estimate Eq.~\eqref{eq:lambda_crit}, we approximate the core sound speed using the virial scaling $c_s^2\sim GM_\star/R_\star$, and the mean ZAMS density,
\begin{equation}
    \bar\rho_\star(M_\star)\sim \rho_\odot\,\frac{M_\star}{\msun}\left(\frac{R_\star}{R_\odot}\right)^{-3},
\end{equation}
where $\rho_\odot =1.4\,{\rm g\,\cm^{-3}}$ is the solar mean density. For low-mass main-sequence stars, we adopt the approximate mass--radius relation $R_\star\propto M_\star^{0.8}$, motivated by empirical and theoretical low-mass stellar mass--radius relations \citep[e.g.,][]{Demory2009,Torres2010}. This gives
\begin{equation}
    c_s(M_\star)=c_{s,\odot}\left({M_\star\over M_\odot}\right)^{0.1},
\end{equation}
where $c_{s,\odot}=690\,{\rm km\,s^{-1}}$ from our solar \texttt{MESA} calibration.

We model the suitable-companion fraction as a relative weight rather than an absolute occurrence rate. For a PBH of initial mass $M_0$, the largest companion separation that permits inspiral within the reference $10\,{\rm Gyr}$ timescale is
\begin{equation}\label{eq:ap}
    a_{p,\max}\approx 5\,{\rm AU}\left({M_0\over10^{22}\,{\rm g}}\right)^{2/3}\left({M_\star\over M_\odot}\right)^{-1/3}.
\end{equation}
We require the companion orbit to lie outside a fiducial $a_{p,\min}=2.5\,R_\star$. Assuming companions are approximately uniform in $\ln a_p$, we model the suitable-companion fraction as proportional to the logarithmic width of the allowed separation interval,
\begin{equation}
    f_{\rm comp}(M_\star,M_0)=\min\left[{
    \ln\left\{\max\left[a_{p,\max}/a_{p,\min},1\right]\right\}
    \over
    \ln\left(5\,{\rm AU}/2.5R_\odot\right)},1\right].
\end{equation}
This gives $f_{\rm comp}=0$ when $a_{p,\max}<a_{p,\min}$ and saturates at unity when the available logarithmic interval is comparable to the solar-normalized reference interval. This prescription should be interpreted as a fiducial relative weighting: it captures the fact that lighter PBHs require much tighter companions, but it is not a calibrated occurrence model. Thus, for $M_0=10^{-16}\,\msun$, solar-mass hosts are strongly suppressed by the delivery requirement, even though an already deposited PBH could grow within a solar main-sequence lifetime.

\subsection{Population results}\label{sec:pbh_demographics}
\begin{figure*}
  \centering
  \includegraphics[width=\textwidth]{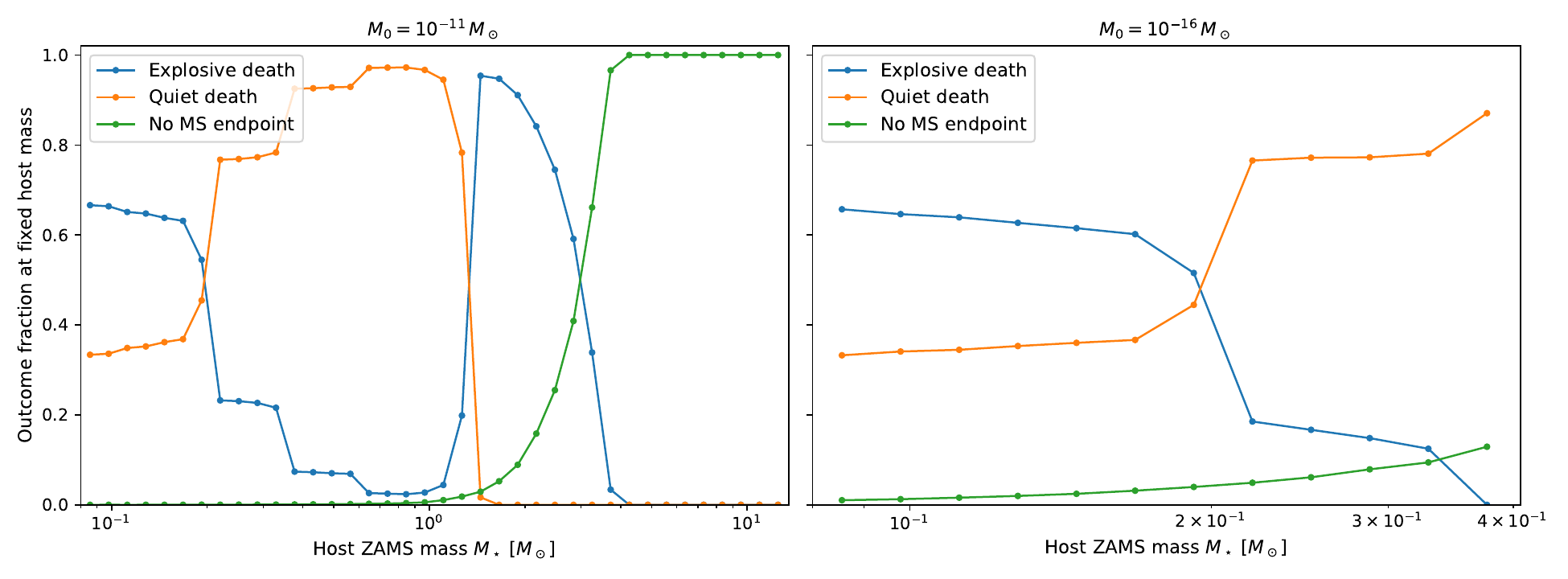}
  \includegraphics[width=\textwidth]{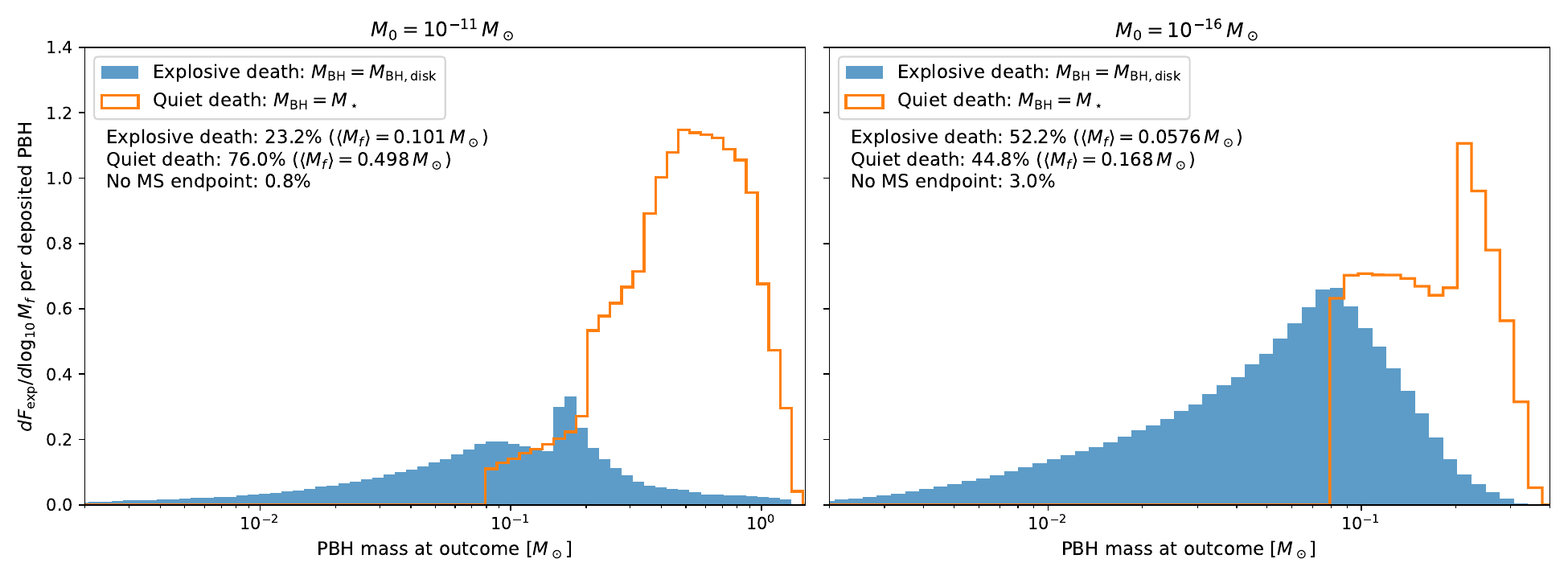}
  \caption{Monte Carlo population outcomes for two initial PBH masses, $M_0=10^{-11}\,\msun$ (left columns), representative of the massive-seed, short-growth-lag limit where $t_{\rm lag}\ll t_{\rm MS}$; and $M_0=10^{-16}\,\msun$ (right columns), representative of the long-lag regime near the minimum seed masses capable of reaching a main-sequence endpoint. Systems are classified as explosive if the disk-formation threshold is reached before the PBH consumes the host, $M_{\rm BH,disk}<M_\star$ and $t_d<t_\star$. Systems for which the PBH consumes the star before an explosive disk forms are classified as quiet deaths, while systems that reach neither endpoint before the end of the modeled main-sequence window are labeled no MS endpoint. {\bf Top panels}: The outcome fractions as a function of host ZAMS mass. The lower-$M_0$ channel is more strongly filtered by growth time, stellar spin-down, and companion-delivery requirements, resulting in a limited host mass range. {\bf Bottom panels}: The capture-weighted PBH mass distribution at the terminal outcome. Explosive deaths are assigned $M_f\approx M_{\rm BH,disk}$ with median $\langle M_f\rangle\lesssim 0.1\,\msun$, while quiet deaths are assigned $M_f\approx M_\star$. The histograms are normalized per deposited PBH, so the area under each component gives the corresponding outcome fraction. Accordingly, the two columns compare conditional outcomes at fixed $M_0$. Their relative cosmic occurrence additionally depends on the PBH mass function and the underlying companion and host-star populations.}
  \label{fig:pbh_pop}
\end{figure*}

We apply the Monte Carlo procedure described above to two representative initial PBH masses, $M_0=10^{-11}\,\msun$ and $M_0=10^{-16}\,\msun$. The former should be interpreted as a representative example of the massive-seed, short-growth-lag regime, for which $t_{\rm lag}\ll t_\star$ for most relevant hosts. The latter illustrates the low-mass-seed, long-growth-lag regime, where PBH growth can be comparable to the stellar lifetime, and stellar spin-down strongly affects the outcome. The resulting outcome fractions and final PBH mass distributions are shown in Figure~\ref{fig:pbh_pop} and summarized in Table~\ref{tab:population_results}. These seed masses are chosen to bracket the post-delivery evolutionary regimes rather than to predict their relative cosmic abundances. All outcome fractions reported below are conditional on successful delivery at the specified $M_0$. A cosmic population prediction would additionally require convolving the intrinsic PBH mass function with the companion-mass and orbital-separation distributions, host-star demographics, and environmental capture rate, which is beyond the scope of the present work.

The main result is that Hawking stars bifurcate into two terminal branches. If the disk-formation threshold is reached before the PBH consumes the host, $M_{\rm BH,disk}<M_\star$, the system enters the explosive branch and the final PBH mass is $M_f\approx M_{\rm BH,disk}$. If instead the disk threshold lies above the available stellar mass, the PBH consumes the host before an explosive disk forms. We classify such systems as quiet deaths, with a terminal mass of order the consumed stellar mass, $M_f\approx M_\star$.

Systems that reach neither endpoint before $t_\star$ are labeled as having no main-sequence endpoint in the present calculation. Low-mass hosts may retain an embedded PBH beyond the present age of the Universe. If the host evolves off the main sequence before either disk formation or quiet consumption, the PBH may be carried into a red-giant or supergiant phase and later into or near the compact remnant, with negligible accretion luminosity as long as the flow remains disk-free. For low- and intermediate-mass hosts, this could produce a PBH--white-dwarf configuration or delayed quiet consumption. This connects to previous studies of PBH encounters with white dwarfs, including PBH-induced heating, carbon ignition, and possible thermonuclear disruption \citep[e.g.,][]{Montero2019}, although an already embedded central PBH is a distinct evolutionary problem. Extending the accretion analysis of \citet{Cantiello2026} to post-main-sequence environments is therefore an interesting direction for future work.

\begin{table}[h!]
\centering
\caption{Fiducial Monte Carlo outcome fractions and median final PBH masses.}
\label{tab:population_results}
\renewcommand{\arraystretch}{2}
\begin{tabular*}{\columnwidth}{@{\extracolsep{\fill}}lccc}
\hline
\shortstack{$M_0$\\$[\msun]$} &
\shortstack{Explosive\\death} &
\shortstack{Quiet\\death} &
\shortstack{No MS\\endpoint} \\
\hline
$10^{-11}$ &
\shortstack{$23.2\%$\\
$\langle{M}_f\rangle=0.101\,\msun$} &
\shortstack{$76.0\%$\\
$\langle{M}_f\rangle=0.498\,\msun$} &
$0.8\%$ \\
$10^{-16}$ &
\shortstack{$52.2\%$\\
$\langle{M}_f\rangle=0.0576\,\msun$} &
\shortstack{$44.8\%$\\
$\langle{M}_f\rangle=0.168\,\msun$} &
$3.0\%$ \\
\hline
\end{tabular*}
\end{table}

The host-mass top panels of Fig.~\ref{fig:pbh_pop} show that the quiet/explosive outcome is controlled by stellar spin-down, the host-mass-dependent disk threshold, and capture selection. The lowest-mass hosts rotate faster, allowing some of them to satisfy $M_{\rm BH,disk}<M_\star$ despite their small masses. By contrast, stars with $0.2\,\msun\lesssim M_\star\lesssim1\,\msun$ rotate more slowly, so disk formation generally requires early capture, reducing their explosive fraction. Near and above the Kraft break, $M_\star\gtrsim1\,\msun$, magnetic braking is suppressed, and most delivered PBHs reach the explosive branch.

The difference between the two seed masses in the bottom panels of Fig.~\ref{fig:pbh_pop} follows from how growth and capture selection sample this host-mass dependence. For $M_0=10^{-11}\,\msun$, the growth lag is short, but capture is uniform over the stellar lifetime; many PBHs are therefore deposited after their hosts have spun down. Since Eq.~\eqref{eq:Mcrit} indicates that $M_{\rm BH,disk}\propto\lambda(t_d)^{-1}$, late deposition raises the disk-formation mass and can move otherwise rapidly growing systems into the quiet branch. Thus, the $10^{-11}\,\msun$ population is not disk-saturated in the lifetime-capture calculation.

For $M_0=10^{-16}\,\msun$, Eq.~\eqref{eq:ap} suggests that delivery is much more restrictive because $a_{p,\max}\propto M_0^{2/3}$. Solar-mass hosts receive little companion-delivery weight: delivering such a light PBH would require a companion orbit smaller than our adopted minimum separation. This selection effect depletes $M_\star\gtrsim0.4\,\msun$ hosts in Fig.~\ref{fig:pbh_pop} for this PBH population and biases the delivered systems toward faster-rotating, low-mass hosts that are more favorable to disk formation. Under our fiducial companion prescription and flat host-mass distribution, the total delivery normalization for $10^{-11}\,\msun$ seeds exceeds that for $10^{-16}\,\msun$ seeds by a factor of 21. This ratio is model-dependent and should not be interpreted as a prediction of their relative cosmic occurrence rates. Hence, the larger explosive fraction in the $10^{-16}\,\msun$ panel is conditional on successful delivery, and does not imply a larger absolute event rate. The fixed-$M_0$ histograms should therefore be interpreted as outcome distributions for delivered PBHs at fixed $M_0$, not as absolute population predictions without specifying the PBH mass function.

Although the explosive branch has a characteristic median final mass $\langle M_f\rangle \approx 0.08\,M_\odot$, the distribution in Fig.~\ref{fig:pbh_pop} has a non-negligible high-mass tail. This tail is especially important for the electromagnetic and gravitational-wave (GW) signatures discussed in \S\ref{sec:signatures}, because the characteristic Bondi accretion rate, and hence the jet power at fixed efficiency, scales as $P\propto\dot \mbh\propto \mbh^2$. Consequently, median-mass estimates provide representative event properties, but the brightest detectable events may be biased toward the upper end of the disk-forming PBH mass distribution.

\section{Accretion disk feedback}\label{sec:feedback}
If the PBH grows to $M_{\rm BH,disk} \lesssim M_\star$, a rotationally supported accretion disk forms and begins dissipating gravitational energy through disk-driven ejecta and through PBH-powered outflows. The relative importance of these two channels depends sensitively on the disk thickness, magnetization, and spin parameter. For a geometrically thin, radiatively efficient disk, the feedback efficiency is given by the Novikov--Thorne value, which depends on the specific energy at the ISCO and ranges from $\eta \approx 6\%$ for a non-spinning BH to $\eta \approx 42\%$ for a maximally spinning BH. In our scenario, however, the accretion flow cannot cool efficiently: it is neither dense enough for neutrino cooling nor optically thin for radiative cooling. The resulting disk is therefore geometrically thick and advection-dominated, in which case the radiative efficiency is substantially reduced, as a large fraction of the dissipated energy is advected into the BH or carried away by winds.

Purely hydrodynamic disk winds may unbind material against the accretion flow, but sustained mass ejection more plausibly requires a strong large-scale poloidal magnetic field. The seed magnetic field in the accreting material originates from stellar dynamo processes, which predominantly generate a randomly oriented toroidal field. As the toroidal field is advected into the forming disk, it drives internal disk dynamos that convert toroidal field into poloidal flux, enabling the development of magnetically driven outflows.

The total two-sided power of the magnetically driven BH outflows is \citep{Blandford1977,Gottlieb2023},
\begin{equation}\label{eq:L}
    P = \eta_a\,\eta_\phi\,\dot{M}_{\rm BH} c^2,
\end{equation}
where $\eta_a$ is the spin-dependent efficiency of a Kerr BH \citep{Lowell2024},
\begin{equation}
    \eta_a \approx 1.063\,a_\ast^4 + 0.395\,a_\ast^2,
\end{equation}
and $\eta_\phi$ characterizes the enhancement due to magnetic flux threading the BH \citep{Tchekhovskoy2011},
\begin{equation}
    \eta_\phi \approx \left( \frac{\phi}{\phi_{\rm MAD}} \right)^2.
\end{equation}
The dimensionless magnetic flux parameter is defined as
\begin{equation}\label{eq:phi}
    \phi \equiv \frac{\Phi}{\sqrt{\dot{M}_{\rm BH}\rg^2\,c}},
\end{equation}
where $\rg = G\mbh/c^2$ is the gravitational radius of the PBH, $\Phi$ is the magnetic flux threading the PBH, and $\phi_{\rm MAD} \approx 50$ represents the maximum flux achievable once the accretion flow enters the MAD state. Equation~\eqref{eq:L} indicates that estimating the jet power requires knowledge of both the BH spin and the magnetic flux.

In \S\ref{sec:disk}, we showed that $ a_\ast \approx 0.8$, independent of stellar properties and PBH mass, corresponding to $ \eta_a \approx 0.7$. Magnetorotational instability (MRI)-driven poloidal magnetic loops are generated in the outer regions of the accretion disk and advected inward. Because the disk confines the magnetic field, the accumulated poloidal flux threading the BH is expected to grow over time. At the moment the disk first forms, the Bondi mass accretion rate is
\begin{equation}\label{eq:Mb}
    \begin{split}
    \dot{M}_B &= \frac{\pi \rho_0 G^2 \mbh^2}{c_s^3}
    \approx 2\times10^{-5}\,\frac{M_\odot}{\rm s}\times\\&
    \left(\frac{\rho_0}{150\,{\rm g\,cm^{-3}}}\right)
    \left(\frac{\mbh}{0.04\,\msun}\right)^{\!2}
    \left(\frac{c_s}{6.9\times10^{7}\,{\rm cm\,s^{-1}}}\right)^{\!-3},
    \end{split}
\end{equation}
where the numerical coefficient is evaluated at the central conditions of our fiducial \texttt{MESA} model at disk formation, and $\mbh\approx M_{\rm BH,disk}$. Outside the PBH and inside the Bondi radius, the density profile of the quasi-spherical inflow follows a free-fall scaling, $\rho \sim r^{-3/2}$, corresponding to a constant mass accretion rate $\dot{M}_{\rm BH} \approx \dot{M}_B$ in the absence of feedback. However, the outflows launched by the disk and BH reduce the inflow rate, producing a time-dependent accretion rate that declines as $\dot{M}_{\rm BH}(t) \propto t^{-\xi}$, with $0 \lesssim \xi \lesssim 1$ depending on how efficiently the jet clears material from around the Bondi radius \citep[see e.g.,][]{Gottlieb2022,Lalakos2025}. As Eq.~\eqref{eq:phi} reads $  \phi \sim \Phi\dot{M}_{\rm BH}^{-1/2}$, a declining accretion rate necessarily leads to an increasing value of $\phi$, even if the magnetic flux $\Phi$ threading the BH remains roughly constant. Consequently, $\phi$ will grow in time until it reaches the MAD limit.

At the moment of disk formation, the circularization radius is expected to lie very close to the ISCO, so the flow initially forms a compact torus of size only a few gravitational radii. In this state, although MRI-driven turbulence can develop rapidly, the disk is too small to generate large-scale poloidal flux bundles capable of coherently threading the BH and powering a sustained jet. The disk must therefore spread viscously to a larger radius, $R_{\rm d} \sim 100\,r_g$, where the characteristic scale of dynamo-generated magnetic loops becomes comparable to the horizon scale and BH-threading flux can accumulate \citep{Chan2026}. The relevant expansion time is the viscous time,
\begin{equation}
t_{\rm visc}(R)\sim \alpha^{-1}\left(\frac{H}{R}\right)^{-2}\left(\frac{R_d}{r_g}\right)^{3/2}\frac{r_g}{c},
\end{equation}
where for the geometrically thick flow expected here, $H/R\sim 0.5$, and taking the viscosity parameter to be $\alpha\sim 10^{-1.5}$, one finds
\begin{equation}\label{eq:tvisc}
t_{\rm visc}(R_d)\approx 50\,\left(\frac{R_d}{100\,r_g}\right)^{3/2}\left(\frac{\mbh}{0.08\,\msun}\right)\,{\rm ms}.
\end{equation}
Therefore, once the disk forms, it can viscously spread to $\sim 100\,r_g$ on a timescale that is effectively instantaneous compared to the subsequent stellar-feedback timescale. If the disk contains a sufficient amount of seed magnetic flux, likely toroidal,  from the free-falling gas, its rapid spreading enables the generation of poloidal magnetic loops on scales large enough to thread the BH and support jet launching. The magnetic loops generated in the outer disk are likewise advected onto the BH on the $t_{\rm visc}$ timescale, increasing the magnetic flux threading the BH. The subsequent transition to the MAD state, however, does not occur at a fixed time: because the production and accumulation of large-scale poloidal flux depend on a stochastic dynamo process and on the polarity history of the field, the time required to reach MAD is itself expected to be stochastic \citep{Jacquemin-Ide2024b,Chan2026}.

Once MAD is achieved, Eq.~\eqref{eq:L} implies that the jet power becomes $P \sim \eta_a \dot{M}_{\rm BH} c^2$. For $a_\ast \approx 0.8$, the spin efficiency is $\eta_a \approx 0.7$, so the BH rotational energy reservoir becomes significant, and BH-powered outflows are expected to dominate over disk-driven winds \citep{McKinney2012,Narayan2012,Bopp2025}. Evaluating Eq.~\eqref{eq:Mb} at the \texttt{MESA} disk-formation conditions and for $\langle M_f\rangle \approx 0.08\,\msun $ gives $\dot{M}_{\rm BH} \approx 8\times10^{-5}\,M_\odot\,{\rm s^{-1}}$, corresponding to a characteristic jet power of $P \lesssim 10^{50}\,{\rm erg\,s^{-1}}$. However, as we find in \S\ref{sec:numerical}, the feedback-regulated accretion rate is lower by $\sim 2$ orders of magnitude, yielding $P \lesssim 10^{48}\,{\rm erg\,s^{-1}}$. 

\section{Explosive death of Hawking stars}\label{sec:explosion} 

Two competing types of outflows operate in this system. As soon as the disk forms, it launches subrelativistic quasi-isotropic outflows from the weakly magnetized disk. Once sufficient poloidal flux has been generated in the disk and advected onto the BH, collimated relativistic jets emerge and propagate more efficiently through the star. Even for very small magnetic efficiencies, $\eta_\phi \ll 1$, the resulting outflow power exceeds the Eddington luminosity by many orders of magnitude, marking the onset of stellar disruption.

Initially, both the jet and disk-driven outflows cannot propagate far, as they are overwhelmed by the ram pressure of the infalling gas within the Bondi radius. Their energy is therefore redistributed quasi-isotropically, driving a nearly spherical accretion shock through the stellar envelope. For an outflow-regulated density profile, $\rho = A_\rho r^{-1}$ \citep{Gottlieb2022}, an outflow with collimation angle $\theta_j$ advances with a dimensionless head velocity \citep{Harrison2018,GottliebNakar2022}
\begin{equation}\label{eq:betaj}
\small
    \frac{v_j(t)}{c} \approx 0.01\left(\frac{0.5}{\theta_j}\right)\left[\frac{\eta_a \eta_\phi \dot{M}_{\rm BH}c^2}{10^{44}\,{\rm erg\,s^{-1}}}\frac{10^{9}\,{\rm g\,cm^{-2}}}{A_\rho}\frac{1\,{\rm s}}{t}\right]^{1/4},
\end{equation}
where the normalizations are chosen to illustrate that even weakly collimated outflows can traverse the stellar envelope on a timescale of order an hour, producing a quasi-spherical explosion. If, however, relativistic jets emerge before this happens, they will break out on a timescale \citep{Harrison2018}
\begin{equation}\label{eq:breakout}
\small
    t_{\rm bo} \approx 30\,\s\,\left[\left(\frac{\eta_a\eta_\phi\dot{M}_{\rm BH}c^2}{10^{48}\,{\rm erg\,s^{-1}}}\right)^{-1}\left(\frac{\theta_j}{10^\circ}\right)^{4}\left(\frac{R_\star}{R_\odot}\right)^{2}\left(\frac{M_\star}{M_\odot}\right)\right]^{\frac{1}{3}}.
\end{equation}
The total energy deposited in the star during the jet propagation is $E_{\rm exp}=P\times t_{\rm bo}\lesssim 10^{49.5}\,\erg$.

As the jet propagates, it inflates a hot cocoon \citep{Bromberg2011} whose energy is comparable to that of the jet itself. Driven by its high thermal pressure, the cocoon expands quasi-isotropically, ultimately unbinding the stellar envelope and disrupting the star within minutes of jet formation. We note that these breakout timescales are considerably shorter than the timescale over which the BH doubles its mass, indicating that no substantial evolution of the BH spin and mass is expected after disk formation \citep{Jacquemin-Ide2024,Lowell2024}.

\section{Numerical simulations of disk-forming stars}\label{sec:numerical}

\subsection{Numerical Setup}

\subsubsection{\texttt{MESA}}\label{sec:mesa}

We use \texttt{MESA} (release r24.08.1) to evolve rotating $1\,M_\odot$, $Z=Z_\odot$ stars from the zero-age main sequence with a PBH placed at their center at star formation, as a hard inner boundary. Our setup is similar to that of \citet{Bellinger2023}, who also studied the evolution of PBH-hosting stars with \texttt{MESA}. The PBH is represented by the inner boundary mass $M_{\rm center}$, which we identify with $\mbh$. The inner boundary radius $R_{\rm center}$ is set to $\rb$ once the Bondi radius exceeds the grid cell size; at earlier times ($\mbh \lesssim 10^{-6}\,M_\odot$), $R_{\rm center}$ is determined by \texttt{MESA}'s minimum cell size and is much larger than $\rb$. This has no effect on the accretion calculation, since the Bondi rate is computed analytically from the central density and sound speed. Initial PBH masses in the range $M_0=10^{-16}$--$10^{-13}\,M_\odot$ are considered. These masses are chosen to bracket the long- and short-growth-lag regimes after successful PBH delivery. As we show in \citet{Cantiello2026}, spherical PBH growth remains rapid and super-exponential across the relevant mass range. Consequently, $M_0=10^{-13}\,\msun$ and larger seeds such as $M_0\sim10^{-10}\,\msun$, all lie in the short-lag regime for the purposes of the stellar evolution modeled here. Increasing the seed mass within this regime primarily shortens an already brief quasi-spherical growth phase, without materially changing the outcome.

The disk-formation transition is reached within a Hubble time for an initial PBH mass $M_0\gtrsim10^{-16}\,\msun$. The models have initial rotation rates $\lambda=0.1$--$0.4$ of the surface Keplerian value, intended to sample the moderate-to-rapidly rotating part of the observed distribution for young low-mass stars \citep[e.g.][]{Gallet:2013}. They include Spruit--Tayler angular-momentum transport, which keeps the main-sequence rotation profiles close to rigid, consistent with asteroseismic constraints on weak internal differential rotation in low-mass stars. We therefore assume rigid rotation when estimating the specific angular momentum at the Bondi radius.

Because the present work is concerned only with the duration of the quasi-spherical phase and with the conditions at the onset of disk formation, we deliberately neglect any radiative or mechanical feedback from the PBH prior to disk formation, consistent with the analytic estimate $\eta\approx 10^{-2}$ derived in \S\ref{sec:bondi} and in our companion paper, \citet{Cantiello2026}. In particular, during the hyper-critical regime prior to disk formation, very weak feedback is expected ($\eta\ll0.01$). This is because most of the emission from the plasma is localized in the high-temperature regions close to the PBH, but these regions are advected onto the horizon on a timescale shorter than the time required for photons to diffuse out of the Bondi sphere \citep{Begelman1979}.  We therefore do not account for any accretion luminosity throughout the \texttt{MESA} evolution, and the PBH grows at the unimpeded Bondi rate following Eq.~\eqref{eq:Mb}.

The procedure executed at each timestep is:
(i) read the local $\rho_{0}$, $c_{s}$, and $\Omega_\ast$ in the innermost stellar cell adjacent to $R_{\rm center}$;
(ii) compute $\rb$ using Eq.~\eqref{eq:Rb} and $\dot{M}_B$ using Eq.~\eqref{eq:Mb};
(iii) update the PBH mass via $\mbh^{n+1}=\mbh^{n}+\dot{M}_B\,\Delta t$, removing the corresponding mass from the innermost envelope cells;
(iv) reset the inner boundary to the new $\mbh^{n+1}$ and $R_{\rm center}=\rb(\mbh^{n+1})$; and
(v) evaluate the disk-formation diagnostic $j_{\rm B}/j_{\rm ISCO}$ using Eq.~\eqref{eq:jB}, and assuming the surface-averaged stellar angular velocity $\Omega_\ast$ and that the rigid-body rotation holds down to $\rb$. This is physically motivated: material outside $\rb$ remains part of the quasi-hydrostatic stellar reservoir, where angular-momentum transport maintains the nearly rigid rotation profile; gas entering the rapid inner flow is therefore assigned the local corotation angular momentum at $\rb$, and the surface rotation rate is unaffected by the growth of the inner boundary. This is confirmed by our models, which include angular momentum transport via rotational instabilities and magnetic torques, and show that $\Omega_{\rm surf}/\Omega_{\rm k}$ remains constant to within $\sim 5\%$ throughout the evolution. This result 
is consistent with asteroseismic observations of low-mass main-sequence stars, which show nearly solid-body rotation profiles \citep{Aerts:2019}.
The disk-formation criterion uses the spin-dependent Kerr ISCO: we self-consistently track the PBH angular momentum $J_{\rm BH}$ accumulated from the accreted material, compute the dimensionless spin $a_\ast$, and evaluate $j_{\rm ISCO}(a_\ast)$ using the \citet{Bardeen1970} formula for the prograde ISCO of a Kerr BH. The PBH spin grows during accretion, shrinking the ISCO and lowering the angular-momentum threshold for disk formation. The run is terminated when $j_{\rm B}\ge j_{\rm ISCO}(a_\ast)$.

\begin{figure}
  \centering
  \includegraphics[width=\columnwidth]{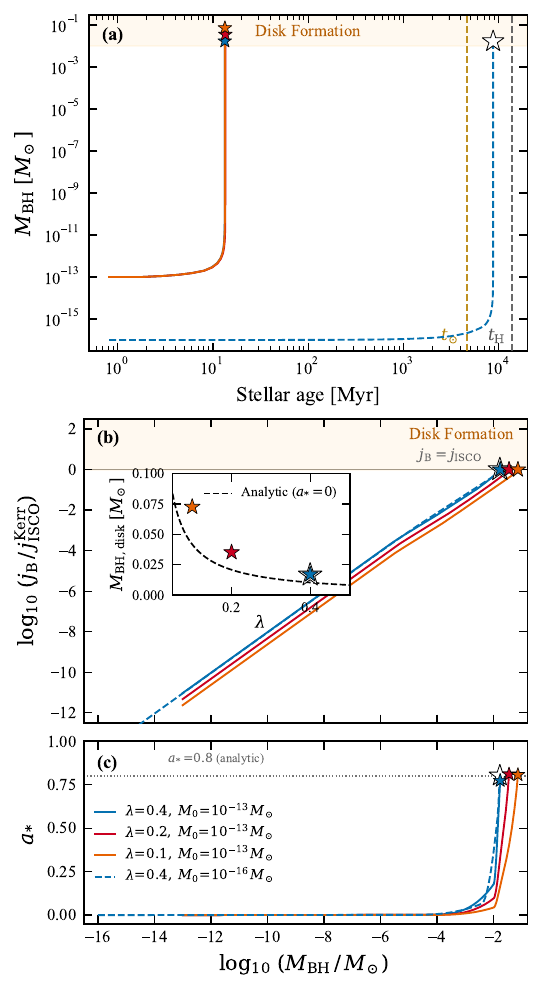}
  \caption{Evolution of the four \texttt{MESA} Hawking-star models toward disk formation. Stars mark disk formation in all three panels. {\bf (a)} PBH mass growth as a function of stellar age. Models with $M_0 = 10^{-13}\,\msun$ reach disk formation in $\sim 70$\,Myr; the $M_0 = 10^{-16}\,\msun$ model requires $\sim 9$\,Gyr. {\bf (b)} Ratio of the specific angular momentum of gas at the Bondi radius to the Kerr ISCO angular momentum, as a function of PBH mass; the steepening near disk formation reflects the self-consistent Kerr correction, as the PBH spins up, the prograde ISCO shrinks, lowering $j_{\rm ISCO}$. \emph{Inset:} disk-formation mass as a function of the dimensionless surface rotation rate $\lambda$, compared to the Schwarzschild ($a_\ast=0$) prediction of Eq.~\eqref{eq:Mcrit} at ZAMS conditions (dashed); the \texttt{MESA} values are systematically higher because stellar evolution increases $c_s$ during the Bondi accretion phase, partly offset by the Kerr correction at $a_\ast\approx 0.8$. {\bf (c)} Dimensionless Kerr spin parameter $a_\ast$ as a function of PBH mass; the spin is negligible during the Bondi phase and rises steeply near disk formation, reaching $a_\ast\approx 0.8$. Panels (b) and (c) share the $M_{\rm BH}$ axis. The horizontal dotted line in (c) marks the analytic prediction $a_\ast = 0.8$ from Eq.~\eqref{eq:a}.}
  \label{fig:mesa}
\end{figure}

Table~\ref{tab:mesa_models} lists our four \texttt{MESA} models, spanning different initial PBH masses and rotation rates, and the resulting PBH mass and spin at disk formation\footnote{The \texttt{MESA} files and output needed to reproduce Table~\ref{tab:mesa_models} and Fig.~\ref{fig:mesa} are available on Zenodo at \href{https://zenodo.org/records/23481715}{doi:10.5281/zenodo.23481715}.}.
Three models share the same initial mass $M_0 = 10^{-13}\,\msun$ but different rotation rates ($\lambda = 0.4$, $0.2$, $0.1$), while a fourth model uses $M_0 = 10^{-16}\,\msun$ at $\lambda = 0.4$ to test independence of the initial PBH mass.

\begin{table}[!ht]
\centering
\caption{\texttt{MESA} models: Initial PBH mass, $M_0$, initial angular momentum in the star relative to Keplerian rotation, $\lambda$, and the PBH mass, $M_{\rm BH,disk}$, and spin, $a_\ast$, at disk formation.}
\label{tab:mesa_models}
\begin{tabular}{cccc}
\hline\hline
$M_0\,[M_\odot]$ & $\lambda$ & $M_{\rm BH,disk}\,[M_\odot]$ & $a_\ast$ \\
\hline
$10^{-13}$ & 0.4 & 0.017 & 0.77 \\
$10^{-13}$ & 0.2 & 0.035 & 0.81 \\
$10^{-13}$ & 0.1 & 0.073 & 0.81 \\
$10^{-16}$ & 0.4 & 0.017 & 0.80 \\
\hline
\end{tabular}
\end{table}

\subsubsection{\texttt{H-AMR}}

Once the shell at the Bondi radius possesses enough angular momentum to circularize outside the PBH’s ISCO, we map the \texttt{MESA} stellar profile into \texttt{H-AMR} \citep{Liska2022} and evolve the subsequent accretion in a 3D GRMHD framework on a fixed
Kerr background, adopting an ideal equation of state with adiabatic index $ \gamma = 5/3 $. We choose the PBH mass at this stage to be $ \mbh \approx 0.04\,\msun $ following our \texttt{MESA} models, which exhibit $M_{\rm BH,disk} = 0.02$--$0.07\,\msun$. Our \texttt{MESA} models self-consistently give $a_\ast = 0.77$--$0.81$ at disk formation, and we adopt a conservative spin of $a_\ast = 0.7$ for the \texttt{H-AMR} simulation. We note that $\mbh/\dot M_B$ is comparable to the global dynamical time of a solar-type star, so the mapped configuration should not be interpreted as a fully self-consistent global stellar equilibrium. Instead, the \texttt{MESA} model supplies the stellar background at disk formation, while the GRMHD calculation follows the local relativistic accretion, magnetic-flux accumulation, and outflow launching.

The inner boundary of the \texttt{MESA} model lies at $r_{\mathsc{MESA},{\rm in}} = 2.2\times 10^{9}\,\cm \approx 3.6\times10^{5}\,\rg$. For radii $r\lesssim 3.6\times10^{5}\,\rg$, we therefore initialize a spherically symmetric density profile by semi-analytically evolving the gas inward under the assumption of free-fall. At the start of our GRMHD simulations, the inner-region profiles at $r<r_{\mathsc{MESA},{\rm in}}$ thus follow the free-fall scalings: a density profile $\rho \sim r^{-3/2}$, a pressure profile $p \sim \rho^{\gamma}$, a radial velocity $v_r \sim r^{-1/2}$, and a constant specific angular momentum. We verify these scalings numerically by evolving the gas with an inner cavity until the solution converges to these profiles. We choose the normalizations such that these quantities match the \texttt{MESA} values at $ r = \rb $. In particular, the specific angular momentum at $ r = R_{\rm B} $ coincides with the angular momentum at the PBH's ISCO at the onset of the GRMHD simulations. For radii $ r > r_{\mathsc{MESA},{\rm in}} $, the \texttt{MESA} models indicate that the gas rotates as a solid body, with $\rho \sim r^{-1/2}$ and $p \sim \rho^\gamma$.

The stellar magnetic field topology predicted by the Tayler-Spruit dynamo \citep{Tayler1973,Spruit2002,Fuller2019a,Skoutnev2024} is predominantly toroidal \citep{Gottlieb2024a}. Our \texttt{MESA} models show a characteristic toroidal field strength of $ B_\varphi \sim 10^6\,{\rm G} $. Therefore, we initialize \texttt{H-AMR} with a large-scale toroidal field of comparable magnitude throughout the grid, extending from the horizon to beyond the Bondi radius.

For the grid setup, we employ spherical polar coordinates $(r, \theta, \varphi)$. The radial direction is logarithmically spaced, extending from slightly inside the event horizon to $6\times10^{9}\,\cm = 10^{6}\,\rg$. Uniform cell sizes are used in both the $\theta$- and $\varphi$-directions. The base grid consists of $N_r \times N_\theta \times N_\varphi$ cells, with $N_r = 288$, $N_\theta = 96$, and $N_\varphi = 96$. To enhance computational efficiency, we employ local adaptive time-stepping together with both static and adaptive mesh refinement. For all simulations, two levels of static refinement are applied in the region $5\,\rg < r < 100\,\rg$ to ensure adequate resolution of the wavelength of the fastest-growing MRI mode in the accretion disk. In addition, we employ adaptive refinement such that, at each radius, one extra refinement level is applied to cells with specific entropy $s > 0.3$.

\subsection{Bondi Accretion and Disk Formation}

Because our models neglect radiative feedback, the stellar structure is almost indistinguishable from a standard rotating main-sequence star throughout the Bondi phase. Figure~\ref{fig:mesa}(b) shows the evolution of $j_{\rm B}/j_{\rm ISCO}^{\rm Kerr}$ as a function of $\mbh$ for the four models. All models reach disk formation (stars), confirming the robustness of the transition. The disk-formation mass scales inversely with the rotation rate, as expected from Eq.~\eqref{eq:Mcrit}: $M_{\rm BH,disk} \approx 0.02$, $0.04$, and $0.07\,\msun$ for $\lambda = 0.4$, $0.2$, and $0.1$, respectively (Table~\ref{tab:mesa_models}). The two models at $\lambda = 0.4$ but with different $M_0$ reach the same $M_{\rm BH,disk}$, confirming that the disk-formation mass is independent of the initial PBH mass. The inset compares the \texttt{MESA} disk-formation masses to the Schwarzschild ($a_\ast = 0$) prediction of Eq.~\eqref{eq:Mcrit}; the \texttt{MESA} values are systematically $\sim 1.5$--$2\times$ higher due to an increase in the central $c_s$ by $\sim 25\%$ during the $\sim 70$\,Myr Bondi growth phase. Disk formation is therefore robust for stars that have not spun down substantially ($\lambda \gtrsim 0.05$--$0.1$), including young systems and the rapid-rotator tail of low-mass stars. For old, slowly rotating stars like the present-day Sun, the PBH would need to consume a significant fraction of the stellar mass before a disk could form, if it forms at all.

Figure~\ref{fig:mesa}(a) depicts the PBH mass growth trajectories, exhibiting a super-exponential growth of $\dot{M}_{\rm BH} \propto M^2$, giving $M(t) = M_0/(1-t/t_{\rm B})$ until the disk-formation criterion is met. The time to disk formation is controlled by the initial PBH mass: $\sim 70$\,Myr for $M_0 = 10^{-13}\,\msun$ and $\sim 9$\,Gyr for $M_0 = 10^{-16}\,M_\odot$. Thus, within the post-capture evolution modeled here, increasing the initial PBH mass shortens rather than lengthens the accretion time. Figure~\ref{fig:mesa}(c) displays the self-consistently computed Kerr spin parameter $a_\ast$, demonstrating a negligible spin during most of the Bondi phase, before rising steeply near disk formation. Remarkably, as Eq.~\eqref{eq:a} predicts, all \texttt{MESA} models exhibit a PBH spin at disk formation of $a_\ast\approx 0.8$, nearly independent of all dimensional quantities. The $\sim 5\%$ residual spin scatter in Table~\ref{tab:mesa_models} reflects the $M^2$-weighted dependence of $J$ on the $(c_s,\Omega_\ast)$ history during the final $t_{\rm B}(M_{\rm BH,disk})\sim$\,weeks of accretion, together with the discrete timestep at which the disk-formation criterion is first satisfied; the analytic value $a_\ast \approx 0.8$ is recovered to within numerical noise in all four models.

\subsection{Disk Formation and Feedback}

\begin{figure}[]
  \centering
  \includegraphics[width=0.45\textwidth]{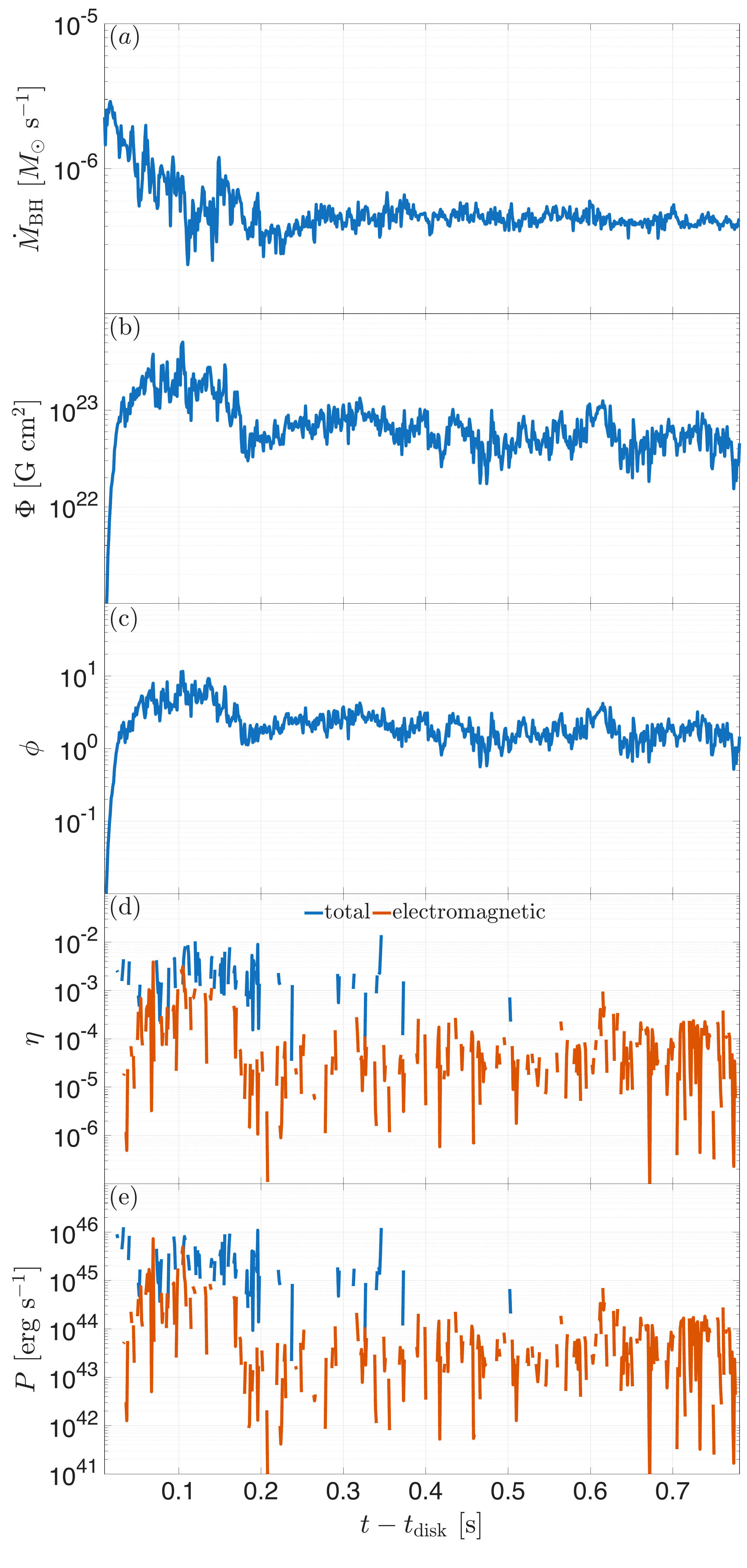}
  \caption{
{\bf (a)} The mass accretion rate initially declines as $\dot{M}_{\rm BH} \sim t^{-0.7}$ due to the formation of the compression shock. Once the flow relaxes, the accretion rate transitions to a plateau, consistent with expectations for freely infalling gas. 
{\bf (b,c)} The magnetic flux, $\Phi$, and dimensionless magnetic flux, $ \phi$, exhibit an early overshoot as toroidal field accumulates in the disk, after which both quantities relax onto a plateau.
{\bf (d)} Hydrodynamic winds launched from the disk tap a fraction of the accretion power and contribute significantly to the total feedback efficiency (blue). Because the disk field is still predominantly toroidal, electromagnetic outflows (red) remain inefficient at this stage.
{\bf (e)} Consequently, the early outflows are dominated by hydrodynamic winds, which exceed the electromagnetic contributions. The segmented appearance of the power curves is a plotting effect: the signed launching power is shown on a logarithmic scale, so intervals in which the net power becomes negative are not displayed.
  }
  \label{fig:BH_profiles}
\end{figure}

\begin{figure*}[]
  \centering
  \includegraphics[width=0.49\textwidth]{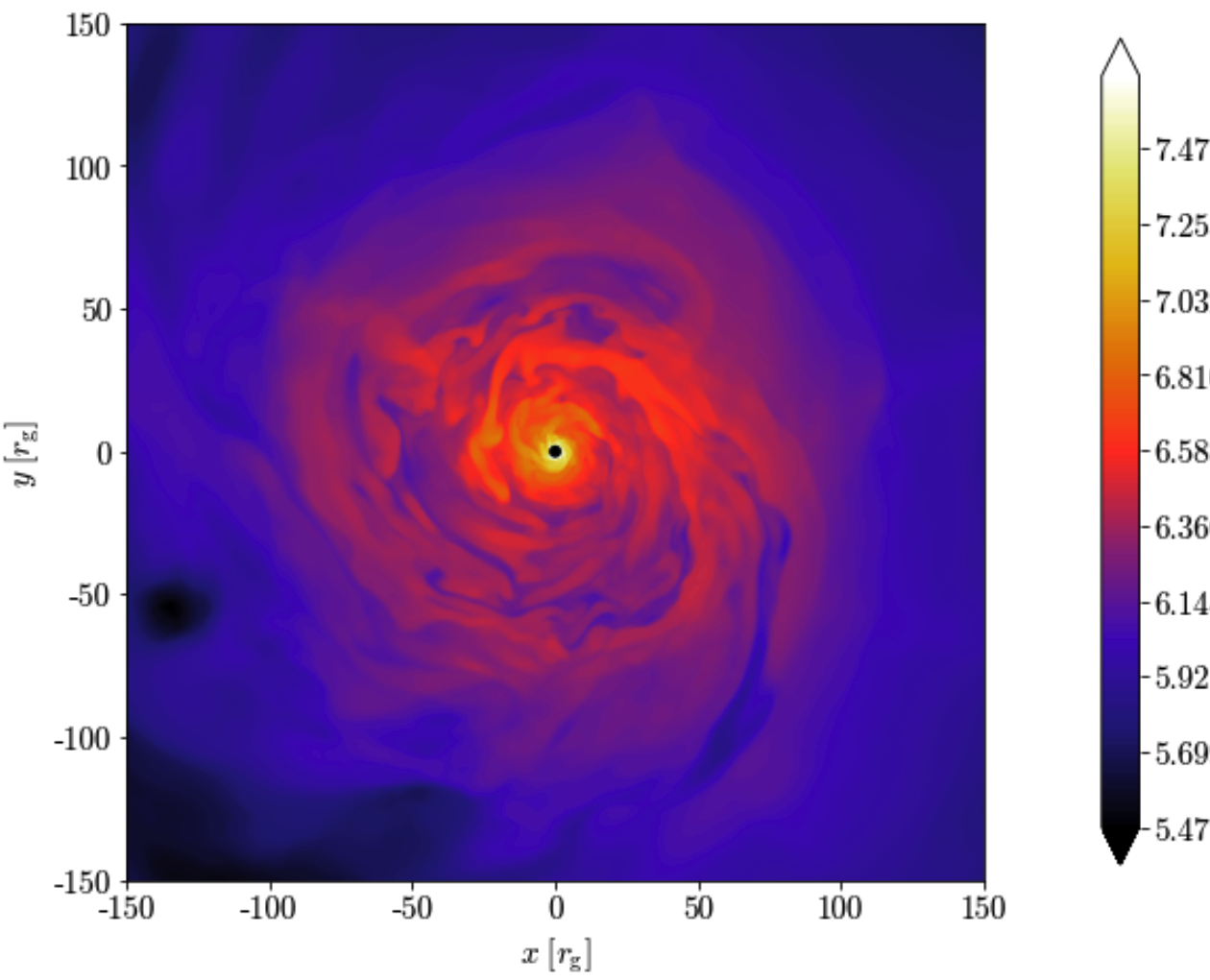}
  \includegraphics[width=0.5\textwidth]{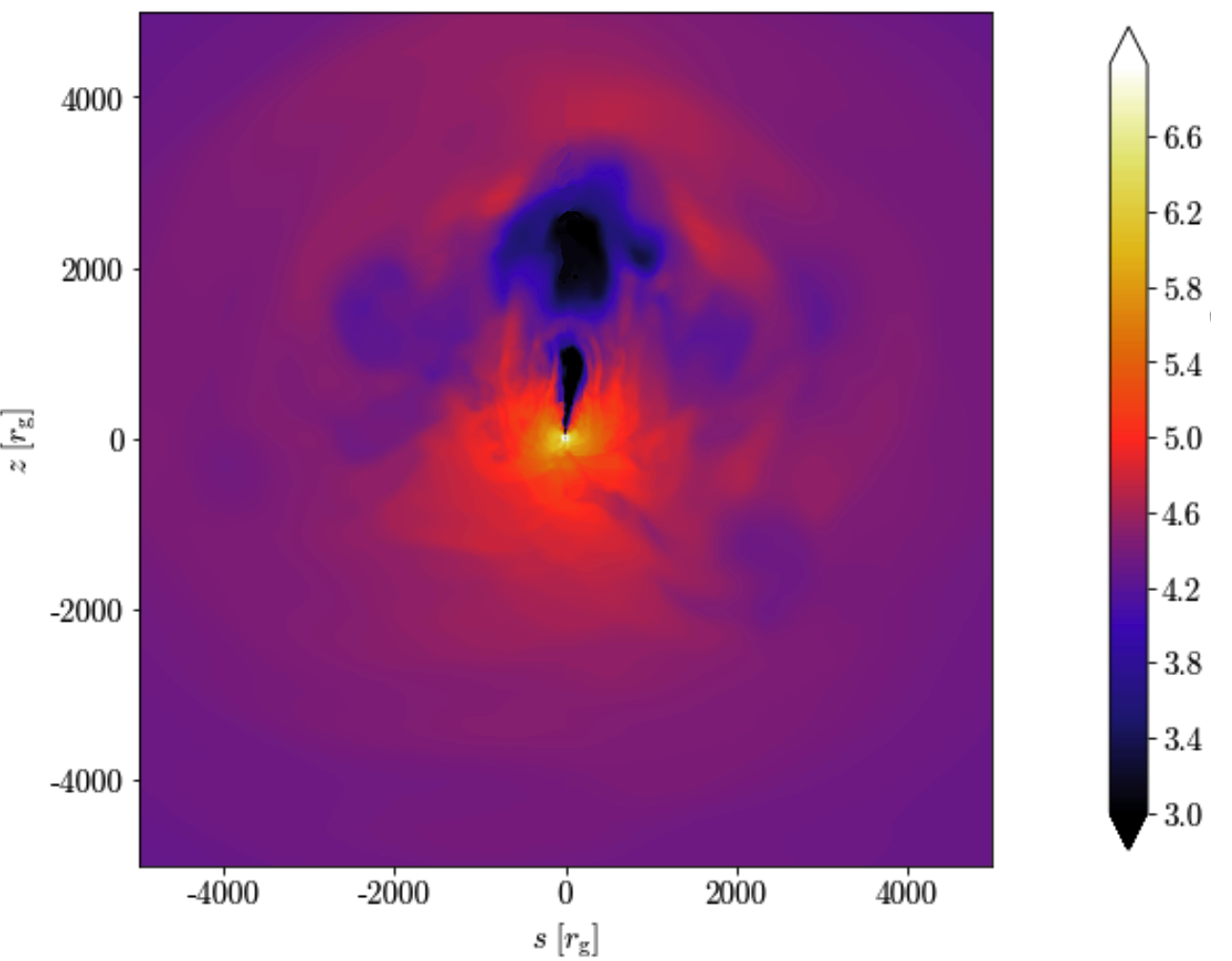}
  \includegraphics[width=0.49\textwidth]{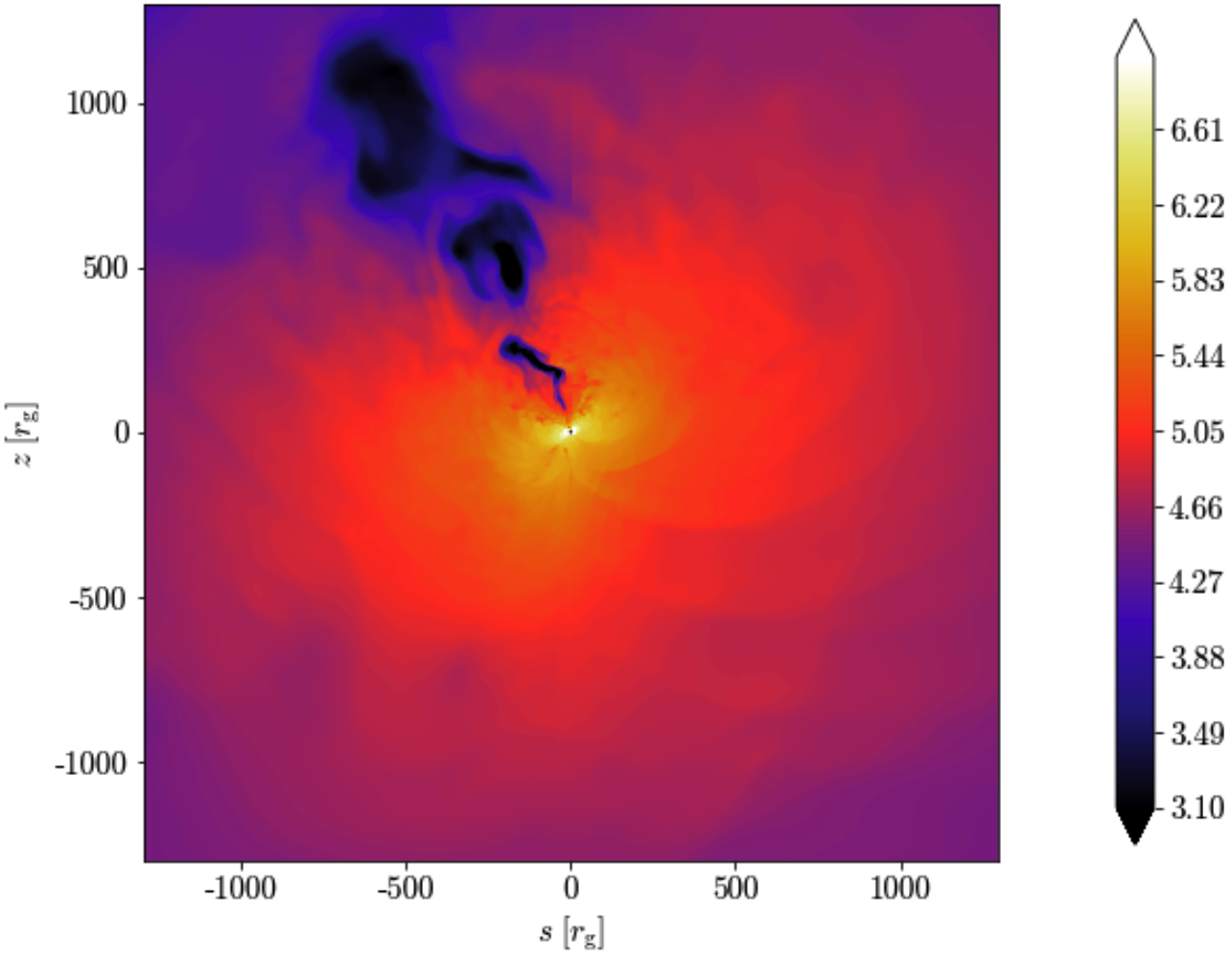}
  \includegraphics[width=0.49\textwidth]{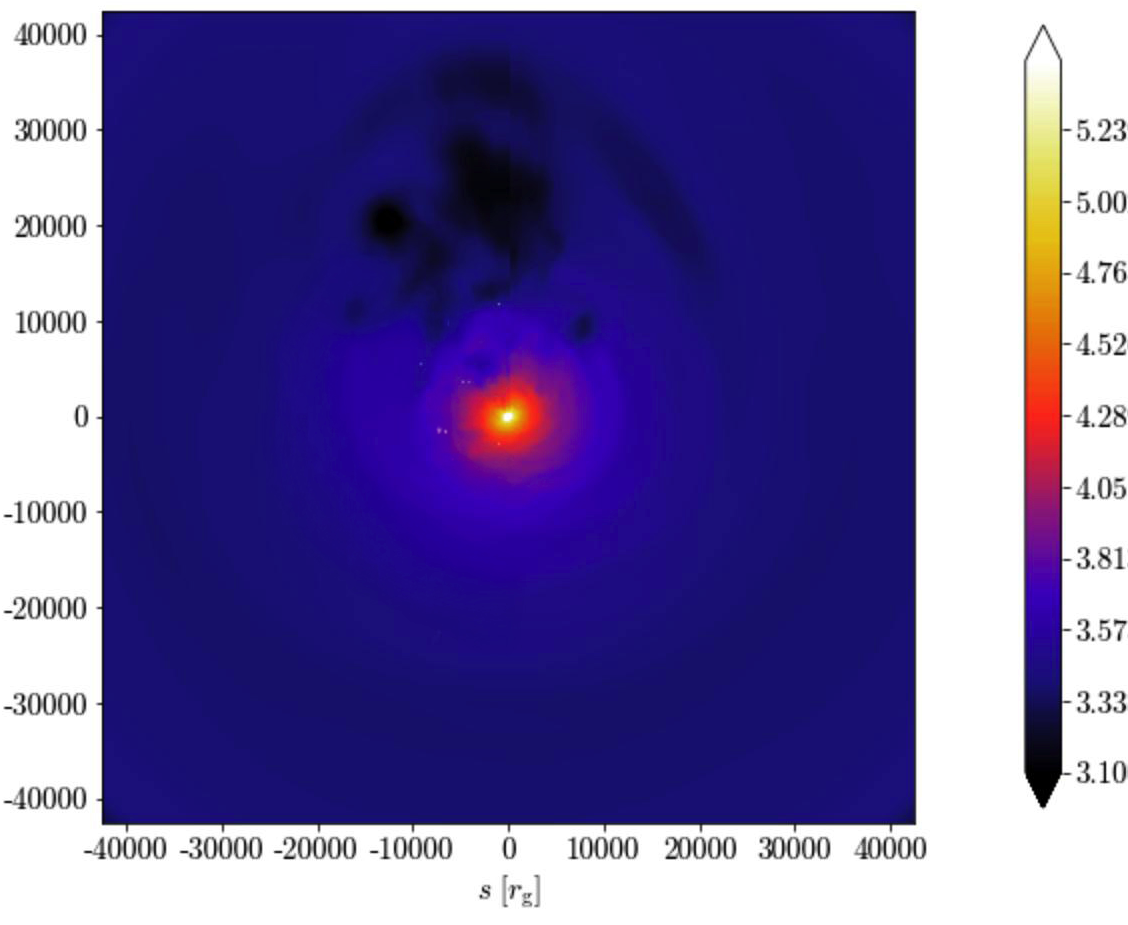}
  \caption{
Planar slices of the logarithmic mass density in $\g\,\cm^{-3}$ at different times. {\bf Top-left}: The equatorial plane at $ t-\td \approx 0.11\,\s $, shortly after the high-density accretion disk forms. {\bf Top-right}: The meridional plane, demonstrating that the disk has accumulated sufficient poloidal magnetic flux to launch a relativistic jet (black) perpendicular to the disk plane. The low-density cavities visible in multiple directions trace the activity of misaligned jets that are subsequently choked by the onslaught of infalling gas. As these outflows are quenched by the ram pressure of the accretion flow, the disk becomes embedded in gas with stochastic angular momentum, leading to a continuous disk tilt. {\bf Bottom-left}: The subsequent launching of a misaligned jet at $ t-\td \approx 0.61\,\s $. {\bf Bottom-right}: A large-scale view from the final snapshot of the simulation at $ t-\td \approx 0.78\,\s \approx 3.8\times 10^6\,\rg/c$, when the jet-driven shock has propagated to $ \sim 4\times 10^4\,\rg\approx 2400\,\km $. 
  }
  \label{fig:maps}
\end{figure*}

At the onset of the GRMHD simulation, the flow develops a transient, outward–propagating compression shock. This occurs because the inner gas begins to collapse inward more rapidly, while the outer layers remain initially more pressure-supported and subsonic. The resulting differential infall compresses the gas at intermediate radii, producing a pile-up that steepens into a shock. As the flow relaxes toward the steady transonic Bondi solution, this shock propagates outward through the infalling material and gradually weakens, after which the standard smooth Bondi profile is recovered.

The expanding reverse shock modifies the ambient medium, establishing a free-fall density profile. For freely infalling gas, such a profile would imply a constant mass accretion rate. However, during the initial phase of the GRMHD simulation, $\sim 0.1\,\s$ following disk formation, the reverse shock suppresses accretion, yielding a steeper decline, $\dot{M}_{\rm BH}\approx 1.3\times 10^{-7}(t/1\,\s)^{-0.7}\,M_\odot\,\s^{-1}$, as demonstrated in Figure~\ref{fig:BH_profiles}(a). Once the shock dissipates and the system relaxes, the mass accretion rate transitions to a plateau consistent with that expected for freely infalling gas inside the Bondi radius.

The top-left panel of Figure~\ref{fig:maps} shows the accretion disk shortly after its formation. Initially, the weak toroidal magnetic field inherited from the stellar interior produces a weakly magnetized disk that is unable to launch electromagnetically driven outflows. As shown in Fig.~\ref{fig:BH_profiles}(b), the magnetic flux remains nearly constant until dynamo action amplifies the field sufficiently on a few $t_{\rm visc}$ timescales. Once enough magnetic flux is threading the BH, it powers a relativistic jet, as illustrated in the top-right panel of Fig.~\ref{fig:maps}. However, the strong ram pressure of the infalling gas disrupts the jet propagation and tilts the accretion disk. As magnetic flux is advected through the disk onto the BH, the resulting jets also become misaligned with the stellar rotation axis and are launched intermittently in varying directions over nearly $4\pi$, as shown in the bottom-left panel of Fig.~\ref{fig:maps} and by the faded, choked jet structures in the top-right panel. The bottom-right panel of Fig.~\ref{fig:maps} shows that despite this disruption, each jet episode deposits energy that drives a shock propagating to large radii. In our simulation, the shock reaches $4\times10^{4}\,\rg \approx 2400\,\km$ at $t-\td = 0.78\,\s$, corresponding to a characteristic shock velocity of $\sim 0.01\,c$, consistent with Eq.~\eqref{eq:betaj}. This implies that, in the absence of a sustained, steady jet that triggers an earlier explosion, the star will unbind and explode within minutes after disk formation.

To assess the time at which a steady jet forms, we find that both the mass accretion rate and the magnetic flux remain approximately constant at early times, similar to the dimensionless magnetic flux shown in Fig.~\ref{fig:BH_profiles}(c), which likewise stays flat. Our results suggest that achieving MAD conditions capable of launching a steady jet, $\phi \sim 50$, may demand a longer dynamo timescale to accumulate the poloidal magnetic flux needed to power strong disk winds and BH-driven outflows. We stress, however, that the disk is only marginally resolved, especially at large radii, so higher-resolution simulations may enable a more efficient dynamo and, as Eq.~\eqref{eq:tvisc} predicts, generate vigorous outflows within the first $\sim 100\,{\rm ms}$ after disk formation. Thus, the simulation should be interpreted as an early-time demonstration of disk-mediated feedback, not as a converged calculation of MRI-driven turbulence, disk-instability cycles, magnetic-flux accumulation, or the final jet/cocoon outcome. If these unresolved processes increase $\phi$, the system moves toward the high-power MAD jet regime estimated below. We defer a detailed study of these processes to future work. Figures~\ref{fig:BH_profiles}(d,e) show the outflow launching efficiency and power, respectively. Because the magnetic field is initially dynamically subdominant in the disk, the early accretion-powered outflows (blue) are launched hydrodynamically rather than electromagnetically (red), resulting in relatively low efficiency. As the magnetic field amplifies, Eq.~\eqref{eq:L} predicts that the launching efficiency can increase substantially, reaching values up to $\sim 45~(69)\%$ for a BH with spin $a_\ast=0.7~(0.8)$ in the MAD state.

As the magnetic flux is confined by the disk, it is not expected to decrease, aside from local stochastic polarity fluctuations. Once dynamo-generated magnetic loops supply enough poloidal flux to the PBH through advection, the total magnetic flux will begin to rise. At the same time, the mass accretion rate is expected to fall as energy deposition into the stellar envelope increasingly exceeds the accretion power. Whichever occurs first --- an increase in magnetic flux or a decline in the accretion rate --- will mark the onset of the growth of $\phi$, ultimately driving the system into the MAD regime, where $\phi \approx 50$. In the MAD state, the jet power will range from $P \sim 10^{45}\,\mathrm{erg\,s^{-1}}\,(\mbh/0.04\,\msun)^2$ if MAD is triggered primarily by a drop in the accretion rate before dynamo amplification becomes significant, up to $P \sim 6\times10^{47}\,\mathrm{erg\,s^{-1}}\,(\mbh/0.04\,\msun)^2$ if the accretion rate remains high and sufficient poloidal flux accumulates on the BH to initiate the MAD state. Eq.~\eqref{eq:breakout} shows that a collimated jet with that power will explode the star within seconds to minutes. The scaling of $P$ with $\mbh$, following Eq.~\eqref{eq:Mb}, is used to calibrate the simulation results to the median PBH mass at disk-forming stars $\langle M_f\rangle=0.08\,\msun$ (\S\ref{sec:pbh_demographics}).

\section{Observational Implications}\label{sec:signatures}
Our results carry important observational implications across both electromagnetic and GW channels. If a PBH has grown sufficiently to form an accretion disk, the associated disk winds and relativistic jets can inject enough energy and momentum into the stellar envelope to unbind the star and power an explosive transient. In this section, we build on the results of the previous sections, adopting $M_{\rm BH}=\langle M_f\rangle=0.08\,M_\odot$ as a representative disk-forming PBH mass, to discuss the transients driven by such events \citep[see][for signatures of the quiet death branch]{Bellinger2023,Caplan2024}, along with their estimated rates and detection prospects. We emphasize, however, that Fig.~\ref{fig:pbh_pop} predicts a distribution of disk-forming PBH masses rather than a single value. Since the jet power scales approximately as $P\propto \dot M_{\rm BH}\propto M_{\rm BH}^2$, the high-mass tail of the explosive population can make a disproportionate contribution to the brightest events and to flux-limited samples. We then turn to the implications for GW sources, before concluding with a discussion of the broader consequences for PBH science.

The properties of the explosive signature are nearly independent of the PBH seed mass $M_0$, but the event rate is not. The transient is set by the PBH mass at disk formation, $M_{\rm BH,disk}\sim10^{-2}$--$1\,\msun$, and by the fixed-point spin $a_\ast\approx0.8$, neither of which retains memory of $M_0$ (\S\ref{sec:disk}); Table~\ref{tab:em_signatures} therefore applies equally to seeds of $10^{-16}\,\msun$ and $10^{-11}\,\msun$. The seed mass enters through the delivery bottleneck of \S\ref{sec:migration}, which imposes three requirements. First, the host must sit in a region of high PBH phase-space density $\rho_{\rm PBH}/\sigma^3$, cf.~Eq.~\eqref{eq:upper_bound}; low-dispersion systems such as dwarf galaxies raise the per-star Liouville bound by $10^{4}$--$10^{5}$ while holding a few percent of the cosmic stellar mass, and plausibly dominate the rate. Second, the host must have a companion massive enough to seed bound, star-crossing orbits by slingshot and close enough for dynamical-friction inspiral to complete within the stellar lifetime, $a_p\lesssim a_{p,\max}\propto M_0^{2/3}$, which falls from $5\,{\rm AU}$ at $\mbh=10^{22}\,{\rm g}$ to $0.4\,{\rm AU}$ at $10^{-13}\,\msun$ and $0.02\,{\rm AU}$ at $10^{-15}\,\msun$, cf.~Eqs.~\eqref{eq:t_df} and \eqref{eq:ap}. Lighter PBHs are not kinematically excluded: they require tighter and correspondingly rarer companion configurations, which are folded into the fractions $f_{\rm comp}$ of Tab.~\ref{tab:rate_table}. Third, conditional on delivery, an explosive rather than quiet outcome favors PBHs whose Bondi growth time is short compared with the spin-down time: at fixed host, a long growth lag $t_{\rm lag}(M_0)$ delays disk formation until $\lambda(t_d)$ has declined, raising $M_{\rm BH,disk}\propto\lambda(t_d)^{-1}$, Eq.~\eqref{eq:disk_success}. Delivery of light seeds partly offsets this by selecting fast-rotating, low-mass hosts (\S\ref{sec:pbh_demographics}, Tab.~\ref{tab:population_results}). The detectable population is therefore weighted toward the upper end of the open asteroid-to-sublunar window, near $M_{\rm BH}^{\rm crit}\sim10^{22}\,{\rm g}$. We do not sharpen this into a predicted mass distribution: that would require convolving the (unknown) PBH mass function with the joint distribution of companion separations and host-star properties, and no census of planetary companions exists far beyond the solar neighborhood. The estimates of Tab.~\ref{tab:rate_table} are crude for this reason.

\subsection{Electromagnetic signatures}\label{sec:EM}

\subsubsection{Emission types}

\begin{table*}[t]
\centering
\small
\caption{Rough order-of-magnitude estimates for electromagnetic signatures of a Hawking-star explosion for a solar-type progenitor.}
\label{tab:em_signatures}
\begin{tabular}{@{}lcccc@{}}
\hline
Signal & Characteristic peak brightness & Timescale & Band & Geometry \\
\hline
Prompt jet emission &
$L_{\gamma,\rm iso}\sim2\times10^{49}\,{\rm erg\,s^{-1}}$ &
$\lesssim$minutes &
soft $\gamma$/hard X-ray &
beamed \\
Shock breakout &
$L_{\rm sbo}\sim5\times10^{45}\,{\rm erg\,s^{-1}}$ &
$\lesssim 1\,{\rm s}$ &
soft $\gamma$/hard X-ray &
weakly beamed \\
Cooling emission &
$L_{\rm pk}\sim2\times 10^{41}\,{\rm erg\,s^{-1}}$ &
$\sim\,1\,{\rm day}$ &
UV/blue optical &
quasi-isotropic \\
Afterglow &
$F_\nu(1\,{\rm Gpc})\sim200\,{\rm \mu Jy}$ &
days &
radio; optical/X-ray if $\theta_{\rm obs}\lesssim\theta_j$ &
beamed evolving to quasi-isotropic\\
\hline
\end{tabular}
\end{table*}

The natural electromagnetic counterparts are a shock breakout signal from the cocoon/disk winds, followed by cooling-envelope emission and synchrotron emission (afterglow) from the expanding ejecta. If a jet successfully escapes the star, additional high-energy emission may be produced shortly after breakout, followed by the jet's own afterglow emission. These signals resemble those of jetted supernovae, but without a $^{56}{\rm Ni}$ decay-powered component. The four main electromagnetic observables are summarized in Table~\ref{tab:em_signatures}.

\paragraph{Prompt emission} Owing to the jet's relatively low power, its propagation may be subject to intense baryon loading of mass $M_{\rm ba}$, which reduces its Lorentz factor at breakout to $ \Gamma_{\rm jet}\sim E_{\rm exp}/M_{ba}c^2$. If $M_{\rm ba}\lesssim 10^{-6}\,\msun$ the jet successfully breaks out of the star while remaining ultra-relativistic, enabling it to produce an X-ray flash or low-luminosity gamma-ray burst (GRB) transient with duration
\begin{equation}
t_\gamma\sim t_{\rm eng}-t_{\rm bo}\sim10^2\,\s,
\end{equation}
where we assume the jet engine (BH activity) time is $t_{\rm eng}\gg t_{\rm bo}$, as accretion may be quenched on this timescale. The true radiated luminosity $L_\gamma\sim \epsilon_\gamma P$, where $\epsilon_\gamma\sim 0.2$ \citep[e.g.,][]{Beniamini2016} is the prompt radiative efficiency. For $P\sim 10^{48}\,\erg\,\s^{-1} $ and $\theta_j\sim0.1$, the isotropic equivalent luminosity is
\begin{equation}
L_{\gamma,\rm iso}\sim \frac{2\epsilon_\gamma P}{\theta_j^2}\sim
2\times 10^{49}\,\frac{\rm erg}{\s},
\end{equation}
where $P$ is the total two-sided jet power.

\paragraph{Shock breakout emission} Independent of whether the jet fully escapes, a quasi-spherical disk- or cocoon-driven shock will eventually reach the stellar surface. If the jet can maintain $ \Gamma_{\rm jet} \gtrsim 3 $, the cocoon is mildly relativistic upon breakout with $\Gamma_{\rm coc} \sim 3$, yielding a signal at temperature \citep{Nakar2012}
\begin{equation}
    T_{\rm sbo} \sim 50\Gamma_{\rm coc}\,{\rm keV}\sim 150\,{\rm keV},
\end{equation}
of duration
\begin{equation}
    t_{\rm sbo} \sim \frac{R_\star}{2c\Gamma_{\rm coc}^2}\sim0.1\,\s,
\end{equation}
possibly followed by a weaker, softer tail extending to $t\gtrsim1\,\s$.
The luminosity released in the breakout layer is
\begin{equation}
L_{\rm sbo}\sim 5\times10^{45} \,\frac{\erg}{\s} \left(\frac{R_\star}{R_\odot}\right)^2\left(\frac{0.1\,\s}{t_{\rm sbo}}\right)\left(\frac{\Gamma_{\rm coc}}{3}\right)^{1.37}.
\end{equation}
If the ejecta is subrelativistic upon breakout, e.g., because the cocoon is less energetic if the jet is choked, the shock breakout emission will be longer, softer, and fainter.

\paragraph{Cooling emission}
After the breakout, the cocoon expands adiabatically, powering a cooling transient \citep{Nakar2017}. For $E_c\approx E_{\rm exp}\sim 10^{49.5}\,\erg$ and $v_c\sim 0.1 c$, the cocoon's mass is $ M_c \sim 2E_c/v_c^2\sim 10^{-2.5}\,\msun $.
The resulting diffusion time and peak luminosity are then
\begin{equation}
t_{\rm optical}\sim
\left(\frac{3\kappa M_c}{4\pi c v_c}\right)^{1/2}
\sim 1\,{\rm day},
\end{equation}
and
\begin{equation}
L_{\rm optical}\sim
\frac{E_cR_\star}{v_ct_{\rm optical}^2}\sim 2\times10^{41}\,\frac{\erg}{\s},
\end{equation}
respectively. Here, $\kappa=0.34\,\cm^2\,\g^{-1}$ is the electron scattering opacity in hydrogen-rich gas. The cooling emission will be released in ultraviolet/blue optical at a characteristic temperature
\begin{equation}
    T\sim\left[\frac{L}{4\pi\sigma_{\rm sb}\left(t_{\rm optical}v_c\right)^2}\right]^{1/4}\sim 2\times 10^4\,{\rm K}.
\end{equation}

\paragraph{Afterglow} Finally, any successful jet and even a choked jet/cocoon should produce a synchrotron afterglow once the ejecta sweep up the circumstellar or interstellar medium. The radio afterglow is estimated as standard synchrotron emission from the forward shock driven by the escaping jet into the ambient medium \citep[e.g.,][]{Sari1998}. For a uniform external density $n_0$, electron index $p_e \approx 2.2 $, and microphysical fractions of the electron energy, $\epsilon_e$ and magnetic energy, $\epsilon_B$, the characteristic synchrotron frequency is
\begin{equation}
\nu_m \sim10^{10} \,{\rm Hz} \left(\frac{2P\theta_j^{-2}t_{\gamma}}{10^{50}\,\erg}\right)^{1/2}\left(\frac{\epsilon_{e}}{0.1}\right)^{2}\left(\frac{\epsilon_{B}}{10^{-2}}\right)^{1/2}\left(\frac{t}{\rm day}\right)^{-3/2},
\end{equation}
and the peak spectral flux for on-axis observers, $\theta_{\rm obs}<\theta_j$, is
\begin{equation}
F_{\nu,{\rm max}} \sim 200\,{\rm \mu Jy} \, \frac{2P\theta_j^{-2}t_{\gamma}}{10^{50}\,\erg}\left(\frac{n_0\epsilon_B}{10^{-2}\,\cm^{-3}}\right)^{1/2}\left(\frac{D}{1\,{\rm Gpc}}\right)^{-2}.
\end{equation}
At GHz frequencies, the radio light curve rises approximately as $F_\nu\propto t^{1/2}$ until $\nu_m$ crosses the observing band, producing peak flux at
\begin{equation}
t_{{\rm pk}}(\nu=8\,{\rm GHz})\sim1\,{\rm day} \left(\frac{2P\theta_j^{-2}t_{\gamma}}{10^{50}\,\erg}\right)^{1/3}\left(\frac{\epsilon_{e}}{0.1}\right)^{4/3}\left(\frac{\epsilon_{B}}{10^{-2}}\right)^{1/3}.
\end{equation}
The same forward shock also produces on-axis optical and X-ray afterglow emission. In the slow-cooling regime, $F_\nu\propto t^{-3(p_e-1)/4}\nu^{-(p_e-1)/2}$ for $\nu_m<\nu<\nu_c$, and $F_\nu\propto t^{-(3p_e-2)/4}\nu^{-p_e/2}$ for $\nu>\nu_c$, where $\nu_c$ is the cooling frequency. If $\theta_{\rm obs}>\theta_j$, the afterglow from a successful jet is initially de-beamed, the peak flux decreases steeply with viewing angle, and the light curve peaks later, roughly at \citep{Gottlieb2019}
\begin{equation}
t_{\rm pk,off}\sim 3\,{\rm days}\left(\frac{2Pt_{\gamma}}{10^{50}\,\erg}\frac{1\,\cm^{-3}}{n_0}\right)^{1/3}\left(\frac{\theta_{\rm obs}-\theta_j}{15^\circ}\right)^2.
\end{equation}
For a choked jet, the relevant late radio signal is instead the more nearly quasi-isotropic cocoon radio remnant.

\subsubsection{Detectability}
To estimate the detection prospects, we consider the volumetric rate estimates of Tab.~\ref{tab:rate_table} and assume that a sizable fraction of them form disks. The relative importance of the different discovery channels is determined not only by their single-event detectability but also by their beaming and survey-selection effects. For a given channel, the detected rate scales schematically as
\begin{equation}
    \dot N_{\rm det} \sim \mathcal{R}_{\rm HS}\,\frac{4\pi}{3}D^3\,f_{\rm sky}\,f_{\rm duty}\,f_{\Omega},
\end{equation}
where $\mathcal{R}_{\rm HS}$ is the volumetric Hawking-star explosion rate, $D$ is the effective horizon, $f_{\rm sky}$ is the survey sky coverage, $f_{\rm duty}$ is its duty cycle, and $f_{\Omega}$ accounts for geometric selection (beaming). For the cooling transient and the late radio remnant, $f_{\Omega}\approx 1$, whereas for successful prompt jet emission $f_{\Omega}\approx\theta_j^2/2$.

Thus, the prompt channel dominates the detected sample only if its horizon exceeds that of the quasi-isotropic optical/UV channel by $\gtrsim f_{\Omega}^{-1/3}\sim4\,(\theta_j/10^\circ)^{-2/3}$. Across the range spanned by Tab.~\ref{tab:rate_table}, the corresponding all-sky quasi-isotropic discovery rate is
\begin{equation}
    \dot N_{\rm iso}\sim \left(10^{-6}-4\times10^{-1}\right) \, {\rm yr^{-1}}
    \left(\frac{D_{\rm iso}}{100\,{\rm Mpc}}\right)^3
    f_{\rm sky}f_{\rm duty}\, ,
\end{equation}
before cadence losses, while the on-axis prompt rate is smaller by the beaming factor above. Hence, if the true rate lies near the optimistic end of Tab.~\ref{tab:rate_table}, the observed population should be dominated by fainter, nearly isotropic channels; near the pessimistic end, the first detections are more likely to come from the largest-volume beamed channel, namely successful on-axis prompt emission. For the $M_0=10^{-16}\,M_\odot$ channel, however, the present-day explosion rate is additionally smeared by the $\sim$Hubble-time Bondi-growth delay and therefore need not track the instantaneous capture rate.

The prompt jet and shock-breakout signals remain the cleanest high-energy triggers for individual events. The cocoon shock breakout is especially well matched to wide-field hard-X-ray/soft-$\gamma$-ray monitors because it is brief ($\lesssim 0.1$--$1$~s) and hard ($\sim 10^2$~keV), making \textit{Fermi}/GBM and \textit{Swift}/BAT the most natural discovery instruments. \textit{Einstein Probe} provides a soft-X-ray complement: with its 0.5--4~keV Wide-field X-ray Telescope, 3600~deg$^2$ instantaneous field of view, and $\sim$minute alert latency, it is especially well suited to a softer X-ray-flash signal and to any softer, longer-lived breakout tail \citep{Yuan2022,Yuan2025}. A successful jet can provide the largest single-event detection horizon, but its observed rate is reduced by beaming and by the uncertain fraction of systems in which the jet both remains relativistic and successfully escapes the star. Using fiducial prompt horizons
\begin{equation}
    D_{\rm prompt}^{\rm GBM}\sim 5~{\rm Gpc},\qquad
    D_{\rm prompt}^{\rm BAT}\sim 3~{\rm Gpc},\qquad
    D_{\rm prompt}^{\rm WXT}\sim 1.5~{\rm Gpc},
\end{equation}
the corresponding on-axis prompt discovery rates are
\begin{align}
    \dot N_{\rm prompt}^{\rm GBM}
    &\sim \left(10^{-3}-10^{3}\right)
    \left(\frac{\theta_j}{10^\circ}\right)^2
    \ {\rm yr^{-1}},\\
    \dot N_{\rm prompt}^{\rm BAT}
    &\sim \left(10^{-4}-10\right)
    \left(\frac{\theta_j}{10^\circ}\right)^2
    \ {\rm yr^{-1}},\\
    \dot N_{\rm prompt}^{\rm WXT}
    &\sim \left(10^{-5}-1\right)
    \left(\frac{\theta_j}{10^\circ}\right)^2
    \ {\rm yr^{-1}}.
\end{align}

For the weaker cocoon shock breakout, taking fiducial hard-X/$\gamma$ horizons
\begin{equation}
    D_{\rm sbo}^{\rm GBM}\sim D_{\rm sbo}^{\rm BAT}\sim 100~{\rm Mpc},
    \qquad
    D_{\rm sbo}^{\rm WXT}\sim 50~{\rm Mpc},
\end{equation}
one finds
\begin{align}
    \dot N_{\rm sbo}^{\rm GBM}
    &\sim \left(10^{-6}-1\right)
    f_{\Omega,{\rm sbo}}\ {\rm yr^{-1}},\\
    \dot N_{\rm sbo}^{\rm BAT}
    &\sim \left(10^{-7}-10^{-1}\right)
    f_{\Omega,{\rm sbo}}\ {\rm yr^{-1}},\\
    \dot N_{\rm sbo}^{\rm WXT}
    &\sim \left(10^{-8}-10^{-3}\right)
    f_{\Omega,{\rm sbo}}\ {\rm yr^{-1}},
\end{align}
where $f_{\Omega,{\rm sbo}}\lesssim 1$ accounts for the weaker anisotropy of the breakout flash. Even near the optimistic end of Tab.~\ref{tab:rate_table}, blind shock-breakout discoveries are therefore expected to be rare.

If jets are not a common outcome of Hawking stars, at moderate-to-high intrinsic rates the most promising blind-search channel is instead the post-breakout cooling transient. In the fiducial solar-type case considered here, it peaks on a $\lesssim 1\,{\rm day}$ timescale as a fast UV/blue transient. This channel is well matched to \textit{ULTRASAT}, whose near-ultraviolet band, very wide field of view, and high-cadence survey strategy are ideal for an hours-to-day UV flash \citep{BenAmi2022,Shvartzvald2024}. Among current facilities, \textit{ZTF} offers the best wide-field optical search for the nearest and brightest events, with a large instantaneous field of view, 30~s exposures, and near-real-time alert generation \citep{Bellm2019,Masci2019,Patterson2019}. \textit{Rubin} reaches much deeper in a single visit and is therefore especially powerful for such fast blue transients, although its baseline cadence remains suboptimal for transients evolving on $\sim1\,{\rm day}$ timescales, making dedicated high-cadence mini-surveys or targets of opportunity particularly valuable \citep{Ivezic2019}.

Taking fiducial cooling-transient horizons
\begin{equation}
    D_{\rm cool}^{\rm U}\sim 0.15~{\rm Gpc},\qquad
    D_{\rm cool}^{\rm ZTF}\sim 0.05~{\rm Gpc},\qquad
    D_{\rm cool}^{\rm Rubin}\sim 0.4~{\rm Gpc},
\end{equation}
and adopting representative day-scale cadence factors
$f_{\rm cad,U}\approx1$, $f_{\rm cad,ZTF}\approx0.5$, and
$f_{\rm cad,Rubin}\approx0.1$, the corresponding blind discovery rates are
\begin{align}
    \dot N_{\rm cool}^{\rm ZTF} \sim \dot N_{\rm cool}^{\rm U}
    &\sim \left(10^{-8}-10^{-2}\right)\ {\rm yr^{-1}},\\
    \dot N_{\rm cool}^{\rm Rubin}
    &\sim \left(10^{-6}-1\right)\ {\rm yr^{-1}}.
\end{align}

The radio afterglow can be used as a confirmation channel rather than the best blind-discovery channel. For a successful on-axis jet, the GHz afterglow peaks on day timescales with fluxes of order $\sim 100\,\mu{\rm Jy}$ at Gpc distances for the fiducial parameters adopted here, well within targeted follow-up reach of the Karl G.\ Jansky Very Large Array \citep{Perley2011}. MeerKAT is likewise attractive for southern follow-up owing to its high sensitivity and wide field of view \citep{Jonas2016}. For off-axis or choked jets, the radio peak is delayed, and the emission is generally better suited to targeted follow-up than to blind discovery. A simple estimate is
\begin{equation}
    \dot N_{\rm radio,follow}\sim f_{\rm follow}\,\dot N_{\rm trig},
\end{equation}
with $f_{\rm follow}\sim 0.5$ a plausible follow-up completeness factor. Thus, for fiducial on-axis afterglows with $D_{\rm radio}\gtrsim D_{\rm trig}$, the radio confirmation rate should be roughly half the trigger rate.

Taken together, the most promising discovery channel depends on both the intrinsic Hawking-star rate and the uncertain success fraction of relativistic jet launching and breakout. For the fiducial horizons adopted above, an on-axis prompt X-ray/$\gamma$-ray trigger probes the largest effective volume even after the beaming penalty. Therefore, if an $\mathcal{O}(1)$ fraction of Hawking stars launch relativistic jets that remain sufficiently clean of baryons and successfully escape the star, the prompt channel is expected to dominate the detected sample at both the pessimistic and optimistic ends of Tab.~\ref{tab:rate_table}. If, however, successful prompt-producing jets are rare, the situation changes. At the pessimistic end, the expected number of events within the smaller UV/optical horizon is $\ll1$ over typical survey lifetimes, so the first detectable events, if any, are still most likely to come from the largest-volume high-energy channel. At the optimistic end, the quasi-isotropic channels become viable and may dominate the detected population if successful prompt-producing jets are sufficiently rare. In that regime, the $\sim1\,{\rm day}$ UV/blue cooling transient provides the most robust blind-search signature independent of jet breakout and beaming.

We have normalized the estimates above to representative jet powers appropriate for the median disk-forming PBH mass and for uncertain post-disk accretion efficiencies. However, because $P\propto \dot M_{\rm BH}\propto M_{\rm BH}^2$, the high-mass tail of the disk-forming PBH distribution in Fig.~\ref{fig:pbh_pop} would likely power the brightest prompt events. Thus, although our fiducial detectability estimates are median-normalized, flux-limited high-energy samples would preferentially select the most massive disk-forming PBHs, and could include events with a jet power approaching, or even exceeding, those inferred for GRB jets. This is especially plausible because the PBHs in the explosive branch are born with $a_*\approx 0.8$, enabling efficient extraction of rotational energy compared with lower-spin GRB central engines \citep{Gottlieb2023b}.

\subsubsection{Distinguishing features and possible contaminants}

The main contaminants are ordinary core-collapse supernovae, engine-driven supernovae and low-luminosity GRBs, and accreting compact-object binaries. Several observational diagnostics should be able to separate these classes.

\paragraph{Ordinary core-collapse supernovae}
The Hawking-star channel is not tied to the terminal evolution of a massive star. It can therefore occur in older stellar populations and in systems whose progenitors are not stripped-envelope stars or red supergiants. In addition, the predicted ejecta energetics are considerably lower than in ordinary core-collapse supernovae, so the optical transient is expected to be faster and less luminous than a normal supernova. The key qualitative prediction is a fast blue transient with little or no ordinary supernova tail powered by $^{56}{\rm Ni}$ radioactive decay. Finally, prompt hard-X-ray/soft-$\gamma$ emission or an unusually strong radio counterpart relative to the optical luminosity would be atypical for an ordinary core-collapse supernova.

\paragraph{Engine-driven supernovae and low-luminosity GRBs}
A successful Hawking-star jet could look like a low-luminosity GRB or X-ray flash. The main differences are the progenitor class and the accompanying optical event. In the PBH scenario, the progenitor need not be a massive stripped star, and the optical counterpart need not resemble a broad-lined Type Ic supernova. Instead, one expects a much fainter, faster-cooling transient. Thus, a prompt high-energy trigger followed by a fast blue transient without a normal broad-lined supernova, especially in a host environment not dominated by recent massive-star formation, would be a strong discriminator in favor of the Hawking-star channel.

\paragraph{X-ray binaries and microquasars}
An accreting compact-object binary should show a long-lived pre-existing accretion phase: persistent or recurrent X-ray activity, state changes, orbital modulation, ellipsoidal variability, or donor-star radial-velocity signatures. A Hawking star, by contrast, should be comparatively quiet before the terminal event, because the embedded PBH accretes quasi-spherically and remains hidden inside the star until the late disk-forming phase. Deep pre-explosion archival limits and late-time imaging demonstrating that no normal donor survives would favor the PBH interpretation.

In practice, the most compelling Hawking-star candidates would be multi-component events: a sub-second hard-X-ray/soft-$\gamma$-ray breakout flash and/or a minute-long X-ray-flash/low-luminosity GRB signal, followed by a $\sim1\,{\rm day}$ UV/blue cooling transient and then radio afterglow or remnant emission, with no ordinary supernova plateau or $^{56}{\rm Ni}$-powered tail, no pre-existing accreting binary at the same location, and no surviving donor star in late-time imaging. This combination would be difficult to reproduce in standard stellar channels and would strongly motivate a Hawking-star interpretation.

We stress, however, that even a deep non-detection of an ordinary supernova counterpart or of a $^{56}{\rm Ni}$-powered tail would not by itself uniquely establish a Hawking-star origin, since other exotic engine-driven channels may also produce high-energy transients with weak or absent supernova emission. This ambiguity is particularly relevant for the prompt channel: at the optimistic end of Tab.~\ref{tab:rate_table}, the intrinsic Hawking-star explosion rate is already of order $\sim 10^2\,{\rm Gpc^{-3}\,yr^{-1}}$, comparable in order of magnitude to commonly inferred local rates of low-luminosity GRBs. If a substantial fraction of Hawking stars launch successful relativistic jets, some prompt Hawking-star explosions could therefore be hidden within the broader X-ray-flash/low-luminosity GRB population. The observed on-axis prompt rate is of course smaller by the uncertain jet-success fraction and the beaming factor, so this comparison should be understood at the intrinsic-population level. A definitive interpretation will likely require comparison with the rate, environment, multiwavelength behavior, and remnant predictions of the PBH-capture scenario.

\subsection{Gravitational waves}
This scenario also has a possible, although likely rare, GW implication. PBHs have long been discussed as potential subsolar GW sources \citep[see e.g.,][]{Sasaki2018,Chen2020,Nitz2021,Yuan2024,Soni2025,Baumgarte2026,Magaraggia2026}. Our channel differs in that the subsolar object is not simply a primordial relic, but a PBH that has grown inside a star before being left behind as a compact remnant. Such an object is particularly significant because BHs in the subsolar regime are not expected from standard stellar evolution. Any merger with a subsolar BH component would therefore be a strong indicator of a non-standard formation channel, such as a PBH or possibly a low-electron-fraction NS  \citep[e.g.,][]{Metzger2024}.

Our framework makes a specific prediction for the explosive branch. The mass accreted after disk formation is small compared with the BH mass at disk formation, $\dot M_{\rm BH}t_{\rm eng}\ll M_{\rm BH,disk}$, so the asymptotic remnant mass is expected to approximately follow the disk-forming PBH mass distribution in Fig.~\ref{fig:pbh_pop}, with $0.01\,\msun\lesssim M_{\rm BH}\lesssim 1\,\msun$. Likewise, because the BH mass changes little after disk formation, little subsequent spin evolution is expected \citep{Jacquemin-Ide2024}, implying an asymptotic spin $a_\ast\approx 0.8$ for PBHs in the explosive branch. A subsolar compact object with such a high spin would therefore be a distinctive fingerprint of this channel, if found in a merging binary.

To estimate the relevance of this channel for GW searches, we compare the Hawking-star remnant formation rate with the much larger population of stellar-origin compact objects, and then account for the additional requirement that the remnants enter merging binaries. At the optimistic end of our estimates, the formation rate of explosive-branch subsolar remnants is bounded above by the capture rate, $\Gamma_{\rm cap}^{\rm opt}\sim10^2\,{\rm Gpc^{-3}\,yr^{-1}}$, after accounting for the fraction of systems in the relevant remnant-mass range (Tab.~\ref{tab:rate_table}). For comparison, the local core-collapse supernova rate is $\mathcal R_{\rm CCSN}\sim10^5\,{\rm Gpc^{-3}\,yr^{-1}}$. Even if only a conservative fraction $f_{\rm BH}\sim0.1$ of core-collapse events leave BHs, the stellar-origin BH birth rate is still $\sim100$ times larger than the most optimistic Hawking-star subsolar-remnant formation rate.

This comparison of formation rates still significantly overestimates the GW yield. First, massive stellar binaries naturally provide progenitors of compact-object binaries, whereas a Hawking-star remnant must either be retained in a pre-existing compact binary or later assembled into one dynamically. The GW merger rate, therefore, requires an additional binary-evolution factor: the fraction of Hawking-star remnants that end up in compact binaries merging within a Hubble time. Second, subsolar systems have lower chirp masses than the stellar-mass BH binaries that dominate current GW catalogs, reducing their detection horizon and survey volume. Thus, GW detections would be highly diagnostic if found, but should be regarded as a binary-evolution-limited and likely subdominant discovery channel for Hawking stars.

If such binaries form, the most detectable systems are those with remnant masses closest to $\sim1\,\msun$ and compact-object companions, where current and future interferometers such as LIGO/Virgo/KAGRA, Cosmic Explorer, and Einstein Telescope could plausibly identify a subsolar component \citep[e.g.,][]{Evans2021,Branchesi2023,LVK2023,Abac2026}. In this sense, GW observations provide a complementary diagnostic of the remnant population, but the expected rate depends on additional binary-survival and binary-assembly physics beyond the scope of this work.

\subsection{Implications}

Beyond individual transients, this channel may also have implications for the PBH abundance. If capture by stars and the subsequent growth phase are efficient, the rate of these stellar disruptions and the abundance of subsolar BH merger remnants could provide independent observational constraints on the PBH population. Conversely, the absence of such transients or of convincing subsolar merger candidates could constrain the fraction of DM in PBHs. Future work will address potential constraints by providing a detailed understanding of capture rates, stellar demographics, binary survival, and selection effects.

Finally, our analysis of the capture process indicates that stellar capture of light PBHs---those in the mass window where they may constitute all of the DM---is a \emph{rare} occurrence. It requires a binary companion that is sufficiently massive and close to the host star. Given the low occurrence rates of hot Jupiters, stellar capture and subsequent disruption by the PBH will \emph{not} happen for most stars, and will be confined to extreme systems or fortuitous alignments of stars, PBHs, and planets. Therefore, we suspect that distortions of stellar mass functions are always small, and that PBH constraints based on stellar populations~\citep{2025A&A...698A.290E} should be significantly weakened. 

Ultimately, the observable outcome depends sensitively on the later stages of the star disruption, including the post-disk-formation PBH accretion, and the outflow propagation through the star. These processes are not fully resolved in our simulations, and will be studied in detail in follow-up work that will provide quantitative predictions for light curves and spectra. Nevertheless, the possibility of linking PBH capture in stars to unusual stellar explosions and to the formation of subsolar BHs makes this channel a broad phenomenological roadmap for future work.

\section{Conclusions}\label{sec:conclusions}

We have developed a first global framework for the formation, evolution, and observable consequences of Hawking stars --- main-sequence stars that capture PBHs and host them at their centers --- by combining analytic arguments, MESA stellar evolution calculations, 3D GRMHD simulations, and population synthesis. We find that Hawking stars have two possible terminal fates. After capture and migration to the stellar interior, the PBH grows through quasi-spherical, low-efficiency Bondi accretion until either the angular momentum of the captured gas becomes sufficient to form a disk, or the PBH consumes the host before this condition is met. In the first case, disk formation triggers efficient magnetic feedback: disk winds and relativistic jets can disrupt the star, producing an explosive Hawking-star transient powered by a rapidly spinning primordial black hole. In the second case, the system dies quietly, leaving a PBH with mass of order the consumed stellar mass and little or no bright disk-powered emission. The boundary between these branches depends on the PBH seed mass, host mass, capture time, stellar spin-down, and companion architecture. Our conclusions may be summarized as follows:

\begin{enumerate}

    \item \textbf{Two-body stellar capture is inefficient, whereas three-body capture can seed Hawking stars in the open PBH mass window.}
    We find that direct one-pass capture of asteroid-to-sublunar mass PBHs by dynamical friction inside a star is negligibly rare. In contrast, three-body interactions with planetary companions provide a viable channel for capture onto bound, star-crossing orbits, followed by inspiral through repeated dissipative stellar transits. For solar-type stars hosting Jupiter-like companions, this yields a characteristic threshold $M_{\rm BH}^{\rm crit}\sim 10^{22}\,\mathrm{g}$, above which inspiral can occur within a main-sequence lifetime. Stellar systems with tighter companions could capture lighter PBHs.

    \item \textbf{The early embedded growth of the PBH proceeds through a quasi-spherical Bondi-like phase with weak feedback.}
    Once settled at the stellar center, the PBH accretes from the surrounding core in a nearly spherical flow. During this stage, the angular momentum of the gas at the Bondi radius is too small to circularize outside the ISCO, so no disk forms, and neither jets nor dynamically important outflows are produced. The accretion efficiency remains low ($\eta \sim 10^{-2}$, below the thin-disk value assumed in earlier work), allowing the PBH to grow considerably more rapidly than in models that assume sustained, non-negligible radiative feedback throughout the evolution. The principal bottleneck for the event rate is therefore the delivery of PBHs into stars rather than the post-capture growth of the PBH. Once centrally deposited, larger PBH seeds grow more rapidly, while low-mass seeds may require Gyr-to-Hubble-time evolution before disk formation. Translating these conditional outcomes into an absolute cosmic event rate and population distribution therefore requires convolving the intrinsic PBH mass function with the companion-mass and orbital-separation distributions, host-star demographics, and environmental PBH phase-space distribution. These ingredients are not modeled self-consistently here and merit a dedicated population study.

    \item \textbf{Disk formation is the critical transition, but it is conditional on stellar spin and capture age.}
    The transition to efficient feedback is set by the condition that the specific angular momentum of a shell inside the Bondi radius is larger than that at the innermost stable circular orbit. For systems that reach this threshold, the PBH spin at disk formation is close to the fixed-point value $a_\ast\approx0.8$, largely independent of the initial PBH seed mass. The PBH mass at disk formation depends on the host mass, sound speed, density, and especially the stellar rotation rate at the delayed disk-formation time. Stellar spin-down can therefore raise the PBH mass substantially for late captures or low-mass seeds with long growth lags.

    \item \textbf{Hawking stars bifurcate into explosive and quiet terminal branches.}
    If the PBH reaches the disk-formation threshold before consuming the host, the system enters the explosive branch and the final PBH mass is the PBH mass at disk formation. If instead the formal disk threshold lies above the available stellar mass, the PBH consumes the star before an explosive disk forms; this is the quiet-death branch, with terminal mass of order the consumed stellar mass. Monte Carlo population calculations show that both quiet and explosive deaths can represent sizable fractions of the delivered Hawking-star population, with their relative importance controlled by the initial PBH mass, host mass, capture age, stellar spin-down history, and companion-delivery efficiency.

    \item \textbf{When the disk forms, the system rapidly evolves toward strong outflow production and stellar disruption.}
    Although the disk initially forms as a compact torus near the ISCO, it quickly spreads viscously, reaching radii where disk dynamos generate poloidal magnetic loops large enough to coherently thread the PBH. As this flux accumulates, the PBH-disk system is capable of launching jets and strong disk winds. The time required to reach efficient outflow launching is stochastic, as it depends on the dynamo-driven buildup and polarity history of the large-scale poloidal field; nevertheless, the resulting jet power can be estimated as $P\sim 10^{45}$--$10^{50}\,{\rm erg\,s^{-1}}$.
    
    \item \textbf{The explosive branch disrupts the star on dynamical timescales.}
    Our analytic estimates and simulations show that once disk- and PBH-driven outflows are launched, they deposit energy efficiently into the stellar interior. Early on, hydrodynamic disk winds and intermittent, misaligned, partly choked jets inflate a hot cocoon and drive a subrelativistic shock through the envelope. If a relativistic jet successfully emerges, it breaks out within seconds to minutes. Otherwise, the cocoon-driven shock is expected to unbind the star within minutes to about an hour, depending on whether a jet forms. Thus, Hawking stars that reach disk formation do not continue quasi-steadily: the disk phase is the point of no return for explosive disruption.

    \item \textbf{Hawking stars have potentially observable electromagnetic signatures.}
    The electromagnetic counterpart of the explosive branch is expected to be multi-component. Independent of the successful emergence of a jet, a cocoon-driven shock can produce a mildly relativistic breakout flash, potentially in hard X-rays/soft $\gamma$-rays on sub-second timescales, followed by a fast UV/blue cooling transient lasting up to $\sim 1\,{\rm day}$ with $L_{\rm pk}\sim10^{41}$--$10^{42}\,{\rm erg\,s^{-1}}$. If a relativistic jet is launched and successfully breaks out of the star, the event may additionally produce a minute-scale prompt signal resembling a low-luminosity GRB or X-ray flash, followed by a synchrotron afterglow. For the most massive disk-forming PBHs, the jet power may approach or exceed typical GRB jet power owing to the high PBH spin.
    
    The dominant discovery channel depends on the intrinsic Hawking-star rate and on how often relativistic jets are launched and escape. If successful jets are common, prompt high-energy transients may dominate the detected sample and, in optimistic rate estimates, could approach the intrinsic rate of low-luminosity GRBs. If jets are rare, the quasi-isotropic $\sim1\,{\rm day}$ UV/blue transient is the most robust blind-search signature. These explosions do not share the progenitors, timescales, or radioactive tails of core-collapse supernovae. However, electromagnetic observations alone will likely identify strong candidates rather than unique classifications, since other exotic engine-driven channels can also produce weak or absent supernova emission.

    \item \textbf{Hawking-star remnants have potentially observable GW signatures.}
    The explosive branch may leave behind a low-mass, rapidly spinning BH with $a_\ast\approx0.8$ and $0.01\,\msun\lesssim\mbh\lesssim1\,\msun$, while the quiet branch leaves a remnant with mass of order the consumed host. Any future GW detection of a compact binary containing a subsolar or otherwise anomalous low-mass BH would be a striking signature of non-standard compact-object formation. Hawking stars provide a concrete astrophysical route for producing such remnants, with the mass and spin distribution encoding the relative importance of the quiet and explosive branches.
    
    \item \textbf{Hawking stars may provide a new probe of PBHs in the asteroid-to-sublunar mass range.}
    If capture and inspiral occur with sufficient frequency, explosive Hawking-star transients provide the most direct observational probe of PBHs in the mass range $10^{17}$--$10^{23}\,{\rm g}$. Their rates, environments, and electromagnetic signatures could constrain the PBH contribution to dark matter. The surviving PBHs offer a complementary probe through their mass and spin distributions: quiet-consumption systems leave more massive remnants, while explosive systems leave subsolar, high-spin BHs. GW detections of these remnants would be highly diagnostic, but are likely less common because they require an additional binary-retention or binary-assembly channel. The key observable is therefore how the Hawking-star population divides between quiet and explosive fates as a function of PBH seed mass, stellar host mass, capture age, and companion architecture.

\end{enumerate}

Several important questions remain for future study. Nearly every stage of the sequence identified here could form the basis of a dedicated project: PBH capture and delivery in realistic stellar systems; feedback-regulated Bondi growth inside stars, which we study in a companion paper, \citet{Cantiello2026}; population synthesis of PBH demographics; the onset of disk formation; magnetic flux generation and dynamo in PBH disks; jet and wind propagation through stellar envelopes; and the resulting transients and observational electromagnetic and GW signatures. Future population calculations should combine these ingredients with realistic PBH mass functions, companion demographics, and capture-time distributions. In this sense, the present work should be viewed as a roadmap: it identifies the key bottlenecks and bifurcation points that determine whether Hawking stars die quietly, explode electromagnetically, or leave behind low-mass, high-spin compact remnants.

\section*{Software}
\texttt{MESA} \citep{Paxton2011,Paxton2013,Paxton2015,Paxton2018,Paxton2019,Jermyn2023}, \texttt{matplotlib} \citep{4160265}, \texttt{numpy} \citep{5725236}, \texttt{scipy} \citep{2020SciPy-NMeth}, \texttt{H-AMR} \citep{Liska2022}

\section*{Acknowledgments}
We are grateful to Lucy Reading-Ikkanda/Simons Foundation for designing Figure~\ref{fig:sketch}. OG thanks Earl Bellinger and Thomas Baumgarte for detailed feedback. MK and CN thank Yacine Ali-Haimoud, Will East, Andrei Gruzinov, David Hogg, and Andrew MacFadyen for useful discussions. KVT thanks Junwu Huang for helpful feedback. This research used resources of the Argonne Leadership Computing Facility, a U.S. Department of Energy (DOE) Office of Science user facility at Argonne National Laboratory and is based on research supported by the U.S. DOE Office of Science-Advanced Scientific Computing Research Program, under Contract No. DE-AC02-06CH11357 (NeutronStarRemnants project). This research used resources of the National Energy Research Scientific Computing Center (NERSC), a Department of Energy User Facility using NERSC allocation m4603 (award NP-ERCAP0029085). MK and CN are supported by NSF grant PHY-2412899. KVT is supported by the NSF grant PHY-2210551.

\bibliographystyle{aasjournal}
\bibliography{references}

\end{document}